\documentclass[aps,prx,twocolumn,superscriptaddress,nofootinbib,longbibliography]{revtex4-2}

\usepackage{graphicx}
\usepackage{subcaption}
\usepackage{dcolumn}
\usepackage{bm}
\usepackage{xcolor}
\usepackage{braket}
\usepackage{bbold}

\usepackage{amsmath}
\usepackage{hyperref}

\begin{document}

\preprint{AIP/123-QED}

\title{Distant Bell States via Measured Spin Chain Dynamics}

\author{R.J. Costales}\email{ralph.costales.24@ucl.ac.uk}
\affiliation{ 
Department of Physics and Astronomy, University College London, London, United Kingdom
}
\author{S. Bose}%
\affiliation{ 
Department of Physics and Astronomy, University College London, London, United Kingdom
}

\date{\today}

\begin{abstract}

Preparing distant qubits in Bell states is a core requirement for quantum networking and distributed quantum computation. The nonequilibrium unitary dynamics of spin chains have been envisaged as a convenient quantum channel for intra-chip communications. Typically it establishes mixed entangled states between its remote ends, to be purified by entanglement distillation after several uses of the channel to obtain Bell states. Here we show how the establishment of entanglement via spin chain dynamics can be (a) sped up and (b) made to yield a pure entangled state after a single usage of the channel, which is upgradable to a Bell-state without the complexity of purification. To this end, we intersperse unitary evolution for some period with measurements on intermediary spins. We explore two different measurements: a measurement that post-selects for a ground state; and local measurements in the computational basis. We illustrate the mechanism with qubits coupled to the ends of a 1D $XY$ spin chain.

\end{abstract}

\keywords{Suggested keywords}

\maketitle


Quantum computers hold the promise of speeding up the time taken to solve problems considered difficult for classical computers \cite{Shor1994, Grover1997}. A quantum computer must comprise of many qubits that can be individually controlled. Generally, these high control regions (processors) are limited in size, which induced DiVincenzo to include networking as a key requirement to build a scalable quantum computer \cite{divincenzo2000physical}. The size limitation of individual registers could be due to spatial access issues via macroscopic controls \cite{burgarth2007optimal,veldhorst2017silicon,boter2022spiderweb,burkard2023semiconductor} or due to the size of common mode gate mediators \cite{Kielpinski2002}. Quantum computers are thus typically envisaged to be composed of several quantum registers of multiple qubits each \cite{Kielpinski2002} with information between them being exchanged via inter-register channels, dubbed quantum buses \cite{Majer2007}, facilitating large scale quantum computing. Photons are typically considered the optimal channel for quantum communications, though this would require introducing qubit-optics interfaces to one's quantum computing architecture. 

Quantum buses comprising of a line of permanently interacting qubits (essentially a spin chain) are a promising channel for short distance inter-register networking \cite{Bose2003,Bose2007}, capable of quantum state transfer protocols \cite{Christandl2004, Christandl2005, Albanese2004, Karbach2005, burgarth2005conclusive, Xiang2024, Faria2025, Bhattacharjya2024, Monterde2025, Tashkeev2025, banchi2011long}. A quantum bus, a spacer but connector element, would ideally have much less control than a series of repeated processes such as swaps or shuttling require. There may be also energetic advantages of using evolutions through time-independent Hamiltonians (spin chain nonequilibrium dynamics is an example) for quantum state transfer \cite{Lewis2023}. These protocols use non-equilibrium dynamics, so that the spin chain, due to dispersion of excitations, acts as a non-ideal quantum channel that decreases the purity of the state during transmission. Thus it is best to use such channels to transmit
entanglement \cite{Bose2003}, but then the mixed entangled state shared between the ends of the channel will need to be purified and upgraded by entanglement distillation to nearly a Bell-state --a near perfect resource for quantum communications.
Alternatively, mixed entangled states can directly be generated between distant ends of a spin chain from quantum quenches \cite{Wichterich2009exploiting,bayat2010entanglement,estarellas2017robust}, again, it is understood that they have to be purified by entanglement distillation to be useful for high fidelity quantum networking.


\begin{figure}[b!]
    \centering
    \includegraphics[width=\linewidth]{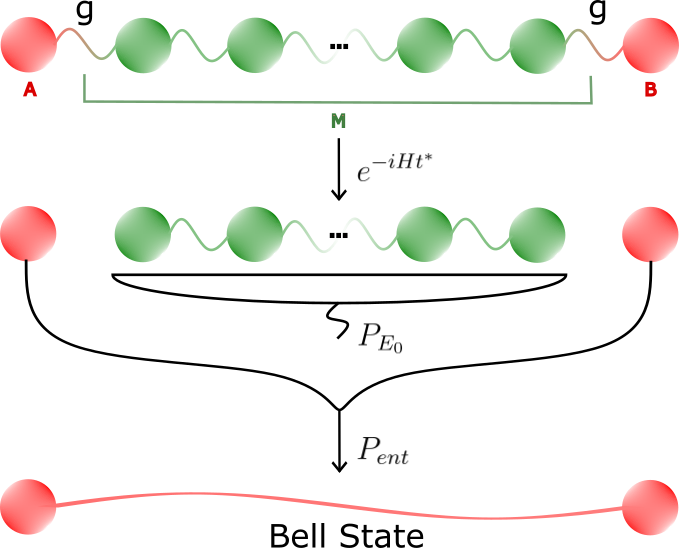}
    \caption{Schematic of the spin chain set-up. The $AB$ qubits are coupled to the end of the bus $M$ by a coupling strength $g$. The spins interact via an $XY$ nearest-neighbour coupling. After a time $t^*$, the $M$ bus spins are projected back onto the ground state of the $XY$ spin chain.}
    \label{fig:schematicsetup}
\end{figure}

Given this scenario of spin chains (without a challenging engineering of each coupling \cite{Christandl2004,Christandl2005,Albanese2004}) typically being imperfect quantum channels between quantum registers, here we ask: Can {\em measurements} on the spin chain channel, after some dynamics, {\em bolster} the entanglement between separated registers while {\em simultaneously also performing purification}? We find an affirmative answer and additionally find that measurements also {\em expedite} the establishment of Bell states between the ends of a spin chain communication channel. Our scheme can be regarded as an application of the ideas of entanglement of assistance \cite{Divincenzo1998}, a quantification of entanglement between two qubits based on their average conversion rate to maximally entangled singlets with the help of a third party, or localizable entanglement \cite{Verstraete2004} to spin chain dynamics for a useful purpose. The scenario considered here is distinct from previous work on measurements, primarily at the ends of spin chains, in dual-rail  \cite{burgarth2005conclusive,vaucher2005arbitrarily,shizume2007quantum}, and other settings \cite{burgarth2007optimal,Feldman2026_PPST, Feldman2026_MQST} for sending encoded qubits. Here the qubits are not encoded, while the measurement is on the body of the spin chain.

The idea presented here is inspired by recent experimental and theoretical developments. On the experimental side, such schemes are on the verge of useful applicability in view of recent realisation of quantum state transfer in 36 qubit 2D superconducting arrays \cite{xiang2024enhanced} following earlier impmentation and capability demonstration work on shorter superconducting and quantum dot spin arrays \cite{li2018perfect,kandel2019coherent,nakajima2018coherent,sigillito2019coherent,farina2025site}. Recently, capability demonstration in organic spin chains \cite{soriano2026exploring} and feasibility of protocols with magnetic atom chains \cite{AcostaCoden2025} have been pointed out, while, for quite some time, similar capabilities have been available for other solid state buses such as chains of Nitrogen spins connecting NV centre spins \cite{yao2011robust}, neutral atom chains in optical lattices \cite{duan2003controlling,banchi2011nonperturbative} and demonstrated for trapped ion chains \cite{jurcevic2014quasiparticle,richerme2014non}, with a recent demonstration in digital ion trap quantum computers \cite{sarkar2026spin}.
Excitonic chains \cite{sgroi2025reinforcement}, Rydberg atom chains \cite{michel2026quantum} and hole-spin chains \cite{sarkar2026all} are also potential candidates, with long range coupled systems also considered recently \cite{avellino2006quantum,lewis2023ion,ahuja2026expediting,Faria2025,marshall2026distinct}.  Such capabilities notwithstanding, there is a crucial {\em asymmetry} between control and measurement: the latter is simply much easier in situations with limited local access. For example, charge states of all quantum dots in an array (which can indicate spin states) can be read out via dispersive coupling to a common circuit \cite{hamonic2025combining}, qubit states in  superconducting arrays can be read similarly by frequency multiplexing and common resonators \cite{bakr2025multiplexed}, while for atomic systems, ions, as well as neutral atoms, such capabilities have been available for a while, using cavities or high resolution fluorescence detection or separated measurement regions \cite{deist2022mid,bluvstein2026fault}, including new capabilities \cite{motlakunta2024preserving}. This begs the question of the current paper: whether measurements on the spin chain bus can be integrated with nonequilibrium dynamics to achieve end to end entanglement across a spin chain bus. Moreover, establishment of entanglement between remote ends, if we are using purely nonequilibrium dynamics, will require a time of the order $\sim N/J$, for an $N$ site spin chain simply because of the speed of excitation propagation in a model with local (such as nearest neighbour) couplings. However, the coherence time of the dynamics may be shorter than that time scale \cite{farina2025site}, which also pushes us to consider the question as to whether measurements can yield the end-to-end entanglement with a smaller duration of nonequilibrium dynamics.

On the theoretical side, there has been an intense focus of late on monitored (measured) systems in many-body condensed matter for new phases and entanglement transitions \cite{fisher2023random,potter2022entanglement,lunt2022quantum, baumer2024efficient, sang2023ultrafast}. Recent work has explored dynamical evolution of an $XY$ chain that mediates the interaction between two distant spins, but a non-Hermitian Hamiltonian arises from continuous measurements rather than a single measurement on the mediator \cite{ahuja2026dynamically}. Similarly, in quantum circuits two qubit entanglement \cite{bao2024finite} and quantum gates \cite{Hoke2023} have been considered using mid circuit measurements.


In our protocol, we explore dynamical evolution followed by one measurement on a spin chain quantum bus. Following the protocol, two initially separable spins which may belong to separate quantum registers interacting indirectly through a quantum bus of spin chains can be probabilistically converted into a Bell state. After a successful run of the protocol, we can thus perform perfect quantum state teleportation algorithms, allowing separate registers to easily collaborate and form a useful quantum computer.

\section{The XY model}

Here we focus on a one-dimensional lattice with open ends and nearest neighbour $XY$ couplings. The $XY$ model is a specific case of the quantum Heisenberg model, a cornerstone of statistical mechanics that is used to study critical points and phase transitions in quantum matter. The $XY$ Hamiltonian for a 1D spin chain with open ends is given by,
\begin{equation}\label{eq:XYmodel}
    H_{XY} = \sum_{i=1}^{N-1} J^x_{i}
    \sigma^{x}_{i} \sigma^{x}_{i+1} + J^y_i\sigma^{y}_{i} \sigma^{y}_{i+1},
\end{equation}
where each $J^{\{x,y\}}_i$ are the strength of the nearest neighbour couplings. This Hamiltonian has SU(2) symmetry, i.e. commutes with all spin operators $[H_{XY},\sigma^a_i]=0$ for $a=x,y,z$. For simplicity, we choose $J^x_i=J^y_i=1$, though this assumption can be relaxed. The Hamiltonian can thus be rewritten as,
\begin{equation}\label{eq:XYmodelalternate}
    H_{XY} = 2\sum_{i=1}^{N-1} \left(
    \sigma^{+}_{i} \sigma^{-}_{i+1} + \sigma^{-}_{i} \sigma^{+}_{i+1}
    \right),
\end{equation}
where $\sigma^{\pm}_i=\frac{1}{2}(\sigma_{i}^{x}+i\sigma^{y}_{i})$. Physically, the $\sigma^{+}_i$ operator transforms a down spin (with respect to the $Z$-basis, or computational basis) into an up spin, and vice versa for $\sigma^{-}_i$. In this form it becomes clear that the number of up and down spins is conserved throughout the dynamical evolution of the spin chain. The $XY$ Hamiltonian can be converted to a free fermion Hamiltonian via Jordan-Wigner transform then diagonalized by a Fourier transform. However, in our application as we will be starting from a ground state (cf. \eqref{eq:InitialState}) and then performing dynamics and further measurement, it was not straightforward to apply this analytic technique, hence numerical exact diagonalisation was used for higher $N>4$ spins.



\section{Entanglement measures}

Crucial to every work that studies the behaviour of entanglement is an entanglement measure of choice. Derived by Lo and Popescu in 2001, one can find the probability that a pure quantum state can be converted into a maximally entangled state \cite{Lo2001}, applied here for the simplest case of converting a two qubit system to a singlet state by local actions and classical communication. The theorem first requires a pure state of interest $\ket{\Psi}$ to be written in the Schmidt basis, i.e.,
\begin{equation}
    \ket{\Psi}=\sum^{N}_{i=1}\sqrt{\lambda_i}\ket{a_i}\ket{b_i},
\end{equation}
where $\{\ket{a_i}\}$ and $\{\ket{b_i}\}$ are orthogonal eigenfunctions and $\sqrt{\lambda_i}$ are ordered Schmidt coefficients such that $\lambda_1\geq\lambda_2$. The Schmidt coefficients can easily be calculated as the square root of the eigenvalues of the reduced density matrix of either the $A$ or $B$ qubit. In the 2 qubit case, the theorem states that there exists a way for $\ket{\Psi}$ to be converted into a maximally entangled state with probability,
\begin{equation}\label{eq:LoPopescu}
    P_{ent}=\operatorname{min} \left\{ 2\lambda_{2},\lambda_1+\lambda_2
    \right\}.
\end{equation}
As $\lambda_1+\lambda_2=1$ due to normalisation and $\lambda_2\leq\frac{1}{2}$ due to the ordering of the Schmidt coefficients, $P_{ent}=2\lambda_2$ for any two qubit pure state.

This measure is centred on the core idea of this work -- gambling entanglement \cite{bennett1996concentrating, gisin1996hidden,bose1999purification}. One can gamble and convert a pure state to one with much higher entanglement. On the flip side, it is possible to lose this gamble with probability $1-P_{ent}$ and end up with an unusable state with less entanglement.

This measure is limited to pure states; in order to compare the entanglement shared between $AB$ where no measurements are made, it is necessary to have a measure that is also valid for mixed states. A popular choice in quantum information theory is logarithmic negativity, defined for bipartite systems as,
\begin{equation}\label{eq:Logneg}
    E_{N}(\rho_{AB})\equiv\operatorname{log_2}
    ||\operatorname{tr}_B(\rho_{AB})||.
\end{equation}
For a two qubit pure state, the logarithmic negativity is $E_N(\ket{\Psi})=\operatorname{log}_2(\lambda_1^2+\lambda^2_2)$. This measure is useful for comparing the logarithmic negativity shared between the mixed state of $AB$ following the dynamical evolution of the total system without measuring the $M$ spins with the purified state of $AB$ following a projective measurement of the $M$ spins where the outcome is recorded. In the event one successfully produces a Bell state, the logarithmic negativity between the two qubits becomes $1$, similarly to the previous entanglement measure $P_{ent}$, thus it is fair to compare the pure state measure \eqref{eq:LoPopescu} with the mixed state measure \eqref{eq:Logneg}. The pure state measure is most practical for analysis as it provides the most useful figure of merit as the probability to obtain a perfect Bell state.

\section{Ground State Projection Protocol}

\subsection{Set-up}\label{ssec: Set Up}


In this section we suggest a theoretical protocol we dub the ground state projection protocol (GSPP), $\mathcal{P}_{GS}$, leveraging non-local measurements on the body of a spin chain. Two qubits $A$ and $B$ are in separate quantum registers and connected by a quantum bus comprised of a 1D chain of spins with $XY$ nearest neighbour couplings -- see Fig. \ref{fig:schematicsetup}. The $AB$ qubits begin in some initial state $\ket{\varphi_0}_{AB}$. The $N_M$ mediating bus qubits are initialised to the ground state of the $XY$ model, $\ket{E_0}_{M}$, where the $M$ subscript denotes the bus spins that mediate the interaction between $A$ and $B$. The combined initial state is thus,
\begin{equation}\label{eq:InitialState}
    \ket{\psi(t=0)}_{MAB} = \ket{E_0}_M \ket{\varphi_0}_{AB}.
\end{equation}

At time $t=0$, the qubits are assumed to be instantaneously coupled to the ends of the mediating spin chain by an $XY$ interaction. The couplings between the bus qubits are taken here to be uniform and set to $1$, while the couplings between the ends of the bus and the register qubits have strength $g$. The combined bus-register system evolves according to the Hamiltonian,
\begin{multline}\label{eq:TotalHamiltonian}
    H=\sum_{i=1}^{N_M} \left(
    \sigma^{x}_{i} \sigma^{x}_{i+1} + \sigma^{y}_{i} \sigma^{y}_{i+1}
    \right) \\
    +g(\sigma^{x}_{A} \sigma^{x}_{1} + \sigma^{y}_{A} \sigma^{y}_{1} + \sigma^{x}_{B} \sigma^{x}_{N_M} + \sigma^{y}_{B} \sigma^{y}_{N_M}).
\end{multline}
At a time $t^*$, the mediating spins are measured in the energy eigenbasis and, with probability $P_{E_0}$, are projected back onto the $XY$ ground state. Mathematically, this is represented by a projective measurement operator $\Pi_{E_0}$,
\begin{equation}
    \Pi_{E_0} = \left(\ket{E_0}\bra{E_0}\right)_{M} \otimes \mathbb{1}_{AB},
\end{equation}
with $P_{E_0}=\bra{\psi(t^*)}\Pi_{E_0}\ket{\psi(t^*)}$. For completeness, the projective operator for the $n^{th}$ excited states would be $\sum_{\alpha}\ket{E_n,\alpha}\bra{E_n,\alpha}_{M} \otimes \mathbb{1}_{AB}$, where one sums over all eigenstates with eigenvalue $E_n$. After the measurement, the mediating qubits are traced out, leaving the $AB$ qubits in the state,
\begin{equation}
    \mathcal{P}_{GS}(\ket{\varphi_0}) = \operatorname{tr}_M \left(
    \frac{\Pi_{E_0} \ket{\psi(t^*)}\bra{\psi(t^*)}_{MAB} \Pi_{E_0}}{P_{E_0}}
    \right)
\end{equation}
The protocol's overall probability of success is given as
\begin{equation}\label{eq:Prob_Success}
    P_s=P_{E_0}P_{ent},
\end{equation}
or in other words, the product of the probability of projecting onto the ground state with the probability of converting such a state into a Bell state.

\subsection{N=2+2 -- Analytic solution}

An analytic exploration of the system's dynamics is done for $N=2+2$, where an XY spin chain with two sites mediates the interaction between two register qubits coupled on either end. It is difficult to scale this problem beyond this small system size -- while methods exist for procuring an effective Hamiltonian between $A$ and $B$ allowing one to increase the spin chain length dramatically, these fail to capture the physics explored by measurements made on the spins that mediate the interaction between $A$ and $B$. Nonetheless, we explore the protocol here explicitly for a small system size to illuminate the process.

The ground state of the $XY$ model with two spins is one of the maximally entangled Bell states, $\ket{\Psi^-}=\frac{1}{\sqrt{2}}\left(\ket{01}-\ket{10}\right)$. The system unitarily evolves according to the Hamiltonian in \eqref{eq:TotalHamiltonian} from an initial state,
\begin{equation}\label{eq:InitialState2+2}
    \ket{\psi(0)}=\ket{\Psi^-}_M\otimes\ket{\varphi_0}_{AB}.
\end{equation}
As per our protocol, a measurement on the mediator spins is made in the energy eigenbasis of the $XY$ model, which projects them onto one of the energy eigenstates $\{\ket{\Psi^-}, \ket{00}, \ket{11}, \ket{\Psi^+}\}$, where $\ket{\Psi^+}=\frac{1}{\sqrt{2}}\left(\ket{01}+\ket{10}\right)$ is another Bell state. Due to the degeneracy of $XY$ Hamiltonian, the state is projected into some superposition of the $\ket{00}$ and $\ket{11}$ states. Naturally, the system's dynamics is wholly dependent on the choice of $\ket{\varphi_0}$.

If the register qubits are in any superposition of aligned states, i.e. $\ket{\varphi_0} = \alpha\ket{00} + \beta\ket{11}$, at time $t$ the system will evolve to,
\begin{equation}\label{eq:2+2;0011}
    \ket{\psi(t)}=\frac{e^{it}}{\tilde{g}}\left(
    a(t)\ket{\varphi_0}_M\otimes\ket{\Psi^-}_{AB}
    +b(t)\ket{\Psi^-}_{M}
    \otimes\ket{\varphi_0}_{AB}
    \right),
\end{equation}
where $\tilde{g}=\sqrt{1+4g^2}$, $a(t)=-2ig\sin(\tilde{g}t)$ and $b(t)=\tilde{g}\cos(\tilde{g}t)+i\sin(\tilde{g}t)$. In this simple case, the state at time $t$ is in a superposition of two states, one that is identical to the initial state in \eqref{eq:InitialState2+2}, and the other wherein $\ket{\Psi^-}$ swaps from the mediator spins to $AB$ and $\ket{\varphi_0}$ vice versa. If we begin with $A$ and $B$ initially separable (i.e. one of $\alpha$ or $\beta$ be zero), a ground state projection here would project the register spins back onto their initial state with zero entanglement. However, projecting onto one of the first excited states $\ket{00}_M$ or $\ket{11}_M$ would project the $AB$ qubits onto the maximally entangled $\ket{\Psi^-}_{AB}$ state. This projection occurs with probability $1-|\tilde{g}^{-1}b(t)|^2$. This example thus illustrates how the GSPP projects even parity initial states back onto themselves, a behaviour that appears to persist for larger $N$, as evidenced by the numerics.

\begin{figure}[t!]
    \centering
    \includegraphics[height=2.5in]{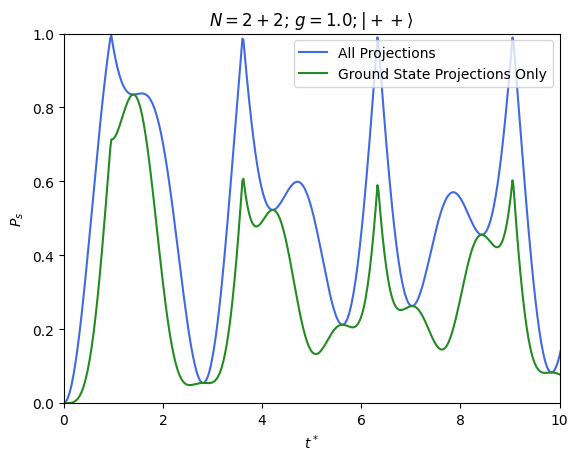}
    \caption{Plot for $P_s$ for a bus-register spin system of size $N=2+2$ with an initial state $\ket{++}$.}
    \label{fig:pMAX Analytics}
\end{figure}

For an initial register state $\ket{\varphi_0}=\ket{++}$ the state evolves as,
\begin{multline}
    \ket{\psi(t)}=\frac{1}{\tilde{g}\sqrt{2}}
    \bigg(
    a(t)\ket{\Phi^+}_M 
    \otimes \ket{\Psi^-}_{AB} \\
    + \ket{\Psi^-}_M \otimes \big(
    b(t)
    \ket{\Phi^+}
    +\tilde{g} e^{it}\ket{\Psi^+}
    \big)_{AB}
    \bigg),
\end{multline}
up to a global phase, where $\tilde{g}$, $a(t)$ and $b(t)$ retain their prior definitions. This appears similar to \eqref{eq:2+2;0011}, taking $\alpha,\beta=\frac{1}{\sqrt{2}}$ such that we obtain the Bell state $\ket{\Phi^+}$ with a probability $\frac{|a(t)|^2}{2\tilde{g}^2}$. However, in this case, a ground state projection now results in the register being projected into a superposition of the $\ket{\Psi^+}$ and $\ket{\Phi^+}$ Bell states:
\begin{equation}
    \mathcal{P}_{GS}(\ket{++})= \frac{ b(t^*)\ket{\Phi^+} + \tilde{g}e^{it}\ket{\Psi^+}}
    {\sqrt{|b(t^*)|^2+\tilde{g}^2}}.
\end{equation}
The probability of the protocol's success $P_s$, as given by \eqref{eq:Prob_Success}, is directly calculated and plotted on Fig. \ref{fig:pMAX Analytics}. However, the Bell state $\ket{\Psi^-}_{AB}$ is also obtained when projecting to one of the higher entangled states. The GSPP represents the lower bound for the probability of successfully creating a Bell state when performing an energy eigenstate measurement on the bus subsystem. This is because projecting the mediating spins onto the higher energy states may also successfully create a Bell pair between $A$ and $B$

\subsection{Numerics for $N>4$}

\begin{figure}[!ht]
    \centering
    \begin{subfigure}[t]{0.49\textwidth}
        \includegraphics[height=2.5in]{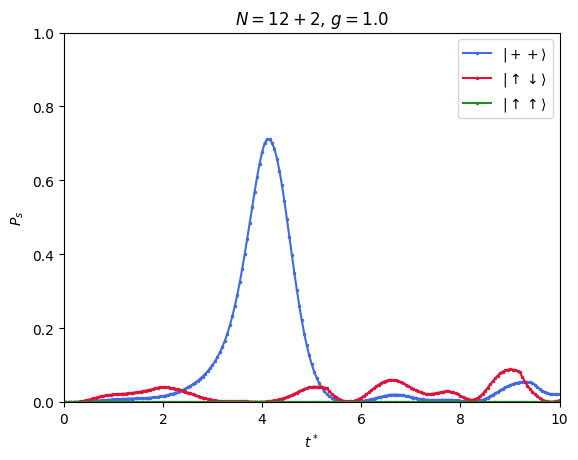}
        \caption{}
         \label{fig:global_pMAX_init}
        \includegraphics[height=2.5in]{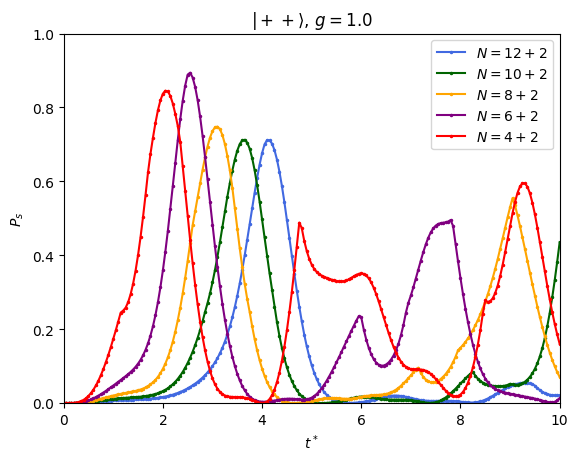}
        \caption{}
        \label{fig:global_pMAX_N}
    \end{subfigure}
    \begin{subfigure}[t]{0.49\textwidth}
        \includegraphics[height=2.5in]{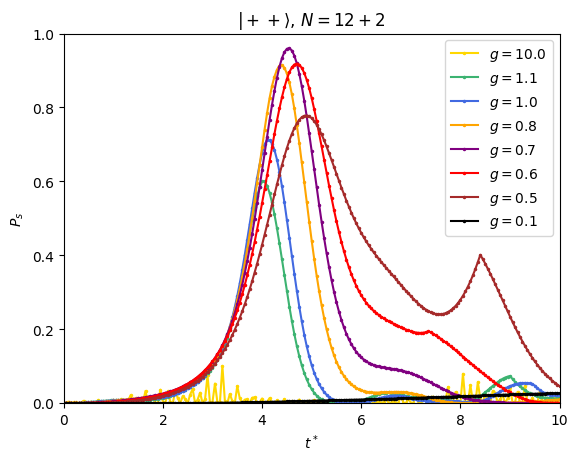}
        \caption{}
        \label{fig:global_pMAX_g}
    \end{subfigure}
    \caption{Plots for $P_s$ made by varying either (a) the initial state $\ket{\varphi_0}$, (b) the number of spins $N$ or (c) the coupling strength $g$ of the $AB$ qubits to the bus, leaving the other variables fixed.}
\end{figure}

To investigate larger system sizes, numerical methods are used to simulate the dynamics of the  bus-register interaction. As the number of energy eigenstates increases, the probability of projecting back onto the ground state dominates. Therefore, for proof of concept, we only calculate the probability $P_{s}$ of successfully obtaining a Bell state after a ground state projection.

For $\ket{\varphi_0} = \alpha\ket{00} + \beta\ket{11}$, just as in the simpler $N=2+2$ case, ground state projections project the $AB$ spins back onto their initial state (see Fig. \ref{fig:global_pMAX_init}). For an initial state $\ket{01}_{AB}$ the probability of projecting onto a Bell state via ground state projection is exceedingly low in short times due to $P_{E_0}$ and $P_{ent}$ being out of phase with respect to each other. Such initial states clearly are a poor choice for the Bell state creation protocol. An initial state $\ket{++}_{AB}$ is far more effective in producing Bell pairs in short time through this protocol, even as system size is scaled up. For $N=12+2$ and $g=1$, the protocol can generate a Bell state with probability $P_s\sim 0.7$ if the mediating spins are projected onto the ground state at $t^*\sim4$.

The effects of varying system size is displayed in Fig.\ref{fig:global_pMAX_N}. The peak of $P_s$ decreases and the time $t^*$ at which this peak occurs increases as $N$ is increased; the GSPP is expected to perform poorly as $N$ is increased, and is best left as a short-distance communication channel. However, the protocol performs best with a system size of $N=6+2$, superior even to the shorter chain of $N=4+2$.

Optimizing the coupling strength $g$ of $A$ and $B$ to the bus appears to greatly improve the efficiency of this protocol. For a slight delay in optimal $t^*$, the optimal $P_s$ nears unity as $g$ is decreased to $\sim0.7$. As the coupling strength $g$ is decreased beyond this optimal coupling strength, $P_s$ decreases again, underlining the importance of careful engineering of the state. Having the $AB$ subsystem be strongly coupled to the mediating spins also appears to poorly affect the efficiency of the protocol overall.

\begin{figure}[b!]
    \centering
    \includegraphics[height=2.5in]{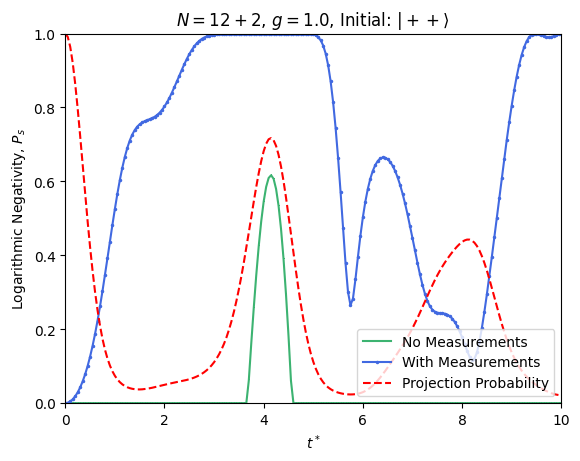}
    \caption{Enhancement of the logarithmic negativity between $A$ and $B$ over time for $g=1$. The green line represents the logarithmic negativity purely through nonequilibrium dynamics. A blue dot represents the logarithmic negativity between $A$ and $B$ after performing a measurement at time $t^*$ following typical dynamical evolution. The probability of projecting onto the ground state by the red dashed line.}
    \label{fig:logneg_g1}
\end{figure}

The entanglement generated between the $A$ and $B$ qubits after the GSPP can be compared to the entanglement generated without measurement by calculating their logarithmic negativity -- see Fig. \ref{fig:logneg_g1}. The first local maximum of the probability of projecting to the ground state coincides with the logarithmic negativity without measurement, around which projections to the ground state in $M$ result in a maximally entangled state in $AB$. Importantly, Fig. \ref{fig:logneg_g1} highlights an interesting aspect of this protocol -- instantaneous collapse of a separate subsystem can convert totally mixed states with no negativity to maximally entangled pure states.

\begin{figure}[b!]
    \centering
    \includegraphics[height=2.5in]{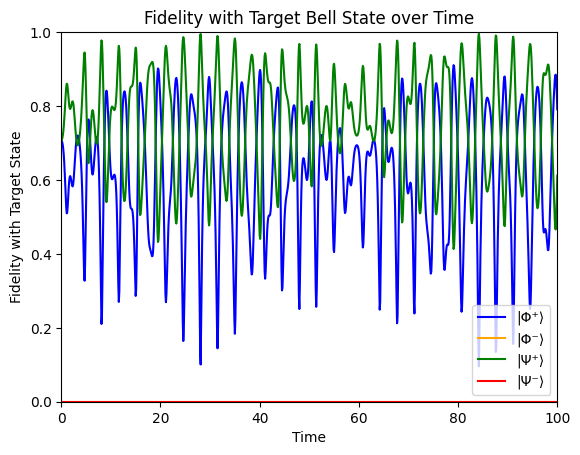}
    \caption{Plot displaying the structure of $\mathcal{P}_{GS}(\ket{++})$ for a system size of $N=10+2$ in the Bell basis for different $t^*$ using the unsquared state fidelity.}
    \label{fig:state_fidelity}
\end{figure}

Fig. \ref{fig:state_fidelity} partially reveals the structure of the subsystem throughout its dynamics after measurement -- attached to the ground state component of $M$ is always $AB$ remaining in superposition of the same two Bell basis states it begins in, $\ket{\Phi^+}$ and $\ket{\Psi^+}$. In other words,
\begin{equation}
    \mathcal{P}_{GS}(\ket{++})=
    \alpha \ket{\Phi^+} + \beta \ket{\Psi^+}
\end{equation}
The coefficients $\alpha$ and $\beta$ evolve according to Fig. \ref{fig:state_fidelity}. Numerical results indicate that any initial state prepared in some equal superposition of any two Bell states will remain in that superposition with varying amplitudes. It follows directly from the fact that we project the mediating spins back to their initial state and the $Z_2$ symmetry (causing $\ket{\uparrow\uparrow
}$ and $\ket{\downarrow\downarrow}$ to maintain the same aplitude) and the mirror symmetry (similarly causing $\ket{\uparrow\downarrow
}$ and $\ket{\downarrow\uparrow}$ to have equal amplitudes) of the spin chain.

These numerical results represent a lower bound to the optimal $P_{s}$, as projections to higher energy states still yield some probability of being converted into Bell pairs. However, there is some utility in requiring a ground state projection specifically, namely that the spin chain may be reused for future iterations of the protocol. These results are promising, and clearly evidences that energy eigenbasis measurements of the mediating spin chain are a reliable tactic to creating Bell pairs at a distance. Although performing such a projective measurement on a spin chain with many sites may be a significant experimental obstacle, some techniques have been suggested \cite{yang2020quantum}.

\section{Local Projection Protocol}

\begin{figure}[b!]
    \centering
    \begin{subfigure}[t]{0.49\textwidth}
        \includegraphics[height=2.5in]{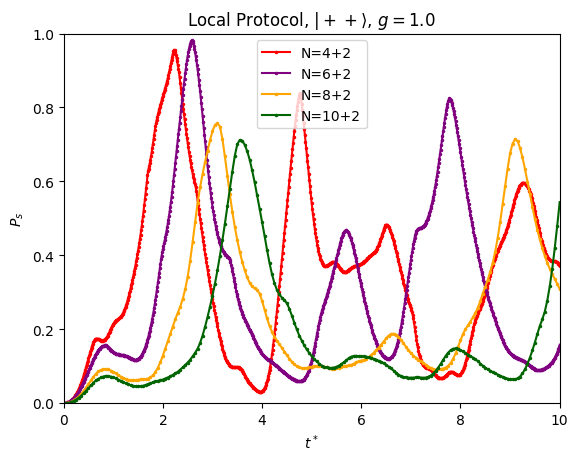}
        \caption{}
        \label{fig:local_pMAX_N}
    \end{subfigure}
    \begin{subfigure}[t]{0.49\textwidth}
        \includegraphics[height=2.5in]{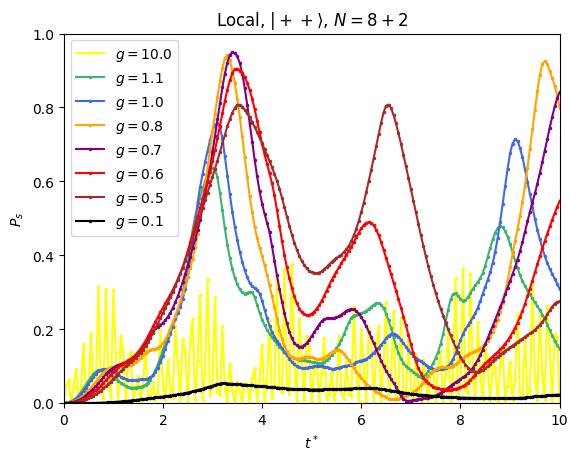}
        \caption{}
        \label{fig:local_pMAX_g}
    \end{subfigure}
    \caption{Plot of $P_s$ for the LPP with $\ket{\varphi_0}=\ket{++}$, varying (a) the number of spins $N$ or (b) the coupling strength $g$ of $A$ and $B$ to the bus, leaving the other variable fixed.}
\end{figure}

Due to the potentially complex nature of a protocol that projects the entire $M$ subsystem onto an energy eigenstate of the $XY$ model, we present an alternative protocol that relies on local computational basis measurements on each individual site of $M$. These measurements would, in principle, be more practical to execute. The local projection protocol (LPP) works in a similar fashion to the GSPP outlined in Sec. \ref{ssec: Set Up}, except now the projection operator is given by,
\begin{equation}
    \Pi_{i} = \left(\ket{i}\bra{i}\right)_{M} \otimes \mathbb{1}_{AB},
\end{equation}
where $\{\ket{i}\}$ are the set of computational basis states, i.e. the eigenstates of the operator $Z^{TOT}$. The measurement outcome must still be recorded before the mediating spins are traced out. The probability of success is now given by,
\begin{equation}
    P_{s}=\sum_i P_i P_{ent, i},
\end{equation}
where $P_{ent,i}$ is the probability of returning a Bell state given a projection $\Pi_i$ has been performed on the system at time $t^*$.

The results still affirm that this protocol is capable of generating Bell states in short time -- see Figures \ref{fig:local_pMAX_N} and \ref{fig:local_pMAX_g}. As in the GSPP, the LPP produces Bell pairs with highest efficiency at a length of $N=6+2$, which sees an increase in the optimal $P_s$ compared to the GSPP (noting that the GSPP results only provide a lower bound to what is possible for an energy eigenbasis measurement, as higher energy states will only bolster the value of $P_s$). The performance of the LPP decreases to match the performance of the GSPP for the longer chains simulated, but it is unclear whether this trend continues to see a decline with increasing system sizes. Like the GSPP, the optimal $g$ for the LPP can be optimized with some engineering of the coupling strengths, allowing for more reliable Bell state creation.

\section{Discussion}

The proposed protocols are capable of generating pure Bell pairs between distant qubits at much faster rates than exclusively nonequilibrium dynamics is capable of. Moving information via teleportation would thus become more efficient. We have studied the regime of up to 12 qubits as efficacy is expected to decrease with longer chains, and the optimal application of linking quantum registers through Bell states would be for short chains. At present, the projective measurements required by us are difficult to perform instantaneously, however, in principle, it should be fully possible to project such disentangled mixed states into pure Bell states with significant probability for success. Future work may involve finding the efficiency of this protocol without requiring the measurements be made instantaneous -- further bridging the gap between the theoretical power of this protocol and its practical limits. It may also be interesting to study this protocol with other spin models or other feasible measurements.


\section{Acknowledgements}

RJC is supported by the Engineering  and Physical Sciences Research Council EP/Y035046/1. SB acknowledges EPSRC-SFI funded project EP/X039889/1 (GeQuantumBus).

\bibliography{aipsamp}

\begin{thebibliography}{71}%
\makeatletter
\providecommand \@ifxundefined [1]{%
 \@ifx{#1\undefined}
}%
\providecommand \@ifnum [1]{%
 \ifnum #1\expandafter \@firstoftwo
 \else \expandafter \@secondoftwo
 \fi
}%
\providecommand \@ifx [1]{%
 \ifx #1\expandafter \@firstoftwo
 \else \expandafter \@secondoftwo
 \fi
}%
\providecommand \natexlab [1]{#1}%
\providecommand \enquote  [1]{``#1''}%
\providecommand \bibnamefont  [1]{#1}%
\providecommand \bibfnamefont [1]{#1}%
\providecommand \citenamefont [1]{#1}%
\providecommand \href@noop [0]{\@secondoftwo}%
\providecommand \href [0]{\begingroup \@sanitize@url \@href}%
\providecommand \@href[1]{\@@startlink{#1}\@@href}%
\providecommand \@@href[1]{\endgroup#1\@@endlink}%
\providecommand \@sanitize@url [0]{\catcode `\\12\catcode `\$12\catcode `\&12\catcode `\#12\catcode `\^12\catcode `\_12\catcode `\%12\relax}%
\providecommand \@@startlink[1]{}%
\providecommand \@@endlink[0]{}%
\providecommand \url  [0]{\begingroup\@sanitize@url \@url }%
\providecommand \@url [1]{\endgroup\@href {#1}{\urlprefix }}%
\providecommand \urlprefix  [0]{URL }%
\providecommand \Eprint [0]{\href }%
\providecommand \doibase [0]{https://doi.org/}%
\providecommand \selectlanguage [0]{\@gobble}%
\providecommand \bibinfo  [0]{\@secondoftwo}%
\providecommand \bibfield  [0]{\@secondoftwo}%
\providecommand \translation [1]{[#1]}%
\providecommand \BibitemOpen [0]{}%
\providecommand \bibitemStop [0]{}%
\providecommand \bibitemNoStop [0]{.\EOS\space}%
\providecommand \EOS [0]{\spacefactor3000\relax}%
\providecommand \BibitemShut  [1]{\csname bibitem#1\endcsname}%
\let\auto@bib@innerbib\@empty
\bibitem [{\citenamefont {Shor}(1994)}]{Shor1994}%
  \BibitemOpen
  \bibfield  {author} {\bibinfo {author} {\bibfnamefont {P.}~\bibnamefont {Shor}},\ }\bibfield  {title} {\bibinfo {title} {Algorithms for quantum computation: discrete logarithms and factoring},\ }in\ \href {https://doi.org/10.1109/SFCS.1994.365700} {\emph {\bibinfo {booktitle} {Proceedings 35th Annual Symposium on Foundations of Computer Science}}}\ (\bibinfo {year} {1994})\ pp.\ \bibinfo {pages} {124--134}\BibitemShut {NoStop}%
\bibitem [{\citenamefont {Grover}(1997)}]{Grover1997}%
  \BibitemOpen
  \bibfield  {author} {\bibinfo {author} {\bibfnamefont {L.~K.}\ \bibnamefont {Grover}},\ }\bibfield  {title} {\bibinfo {title} {Quantum mechanics helps in searching for a needle in a haystack},\ }\href {https://doi.org/10.1103/PhysRevLett.79.325} {\bibfield  {journal} {\bibinfo  {journal} {Phys. Rev. Lett.}\ }\textbf {\bibinfo {volume} {79}},\ \bibinfo {pages} {325} (\bibinfo {year} {1997})}\BibitemShut {NoStop}%
\bibitem [{\citenamefont {DiVincenzo}(2000)}]{divincenzo2000physical}%
  \BibitemOpen
  \bibfield  {author} {\bibinfo {author} {\bibfnamefont {D.~P.}\ \bibnamefont {DiVincenzo}},\ }\bibfield  {title} {\bibinfo {title} {The physical implementation of quantum computation},\ }\href@noop {} {\bibfield  {journal} {\bibinfo  {journal} {Fortschritte der Physik: Progress of Physics}\ }\textbf {\bibinfo {volume} {48}},\ \bibinfo {pages} {771} (\bibinfo {year} {2000})}\BibitemShut {NoStop}%
\bibitem [{\citenamefont {Burgarth}\ \emph {et~al.}(2007)\citenamefont {Burgarth}, \citenamefont {Giovannetti},\ and\ \citenamefont {Bose}}]{burgarth2007optimal}%
  \BibitemOpen
  \bibfield  {author} {\bibinfo {author} {\bibfnamefont {D.}~\bibnamefont {Burgarth}}, \bibinfo {author} {\bibfnamefont {V.}~\bibnamefont {Giovannetti}},\ and\ \bibinfo {author} {\bibfnamefont {S.}~\bibnamefont {Bose}},\ }\bibfield  {title} {\bibinfo {title} {Optimal quantum-chain communication by end gates},\ }\href@noop {} {\bibfield  {journal} {\bibinfo  {journal} {Physical Review A}\ }\textbf {\bibinfo {volume} {75}},\ \bibinfo {pages} {062327} (\bibinfo {year} {2007})}\BibitemShut {NoStop}%
\bibitem [{\citenamefont {Veldhorst}\ \emph {et~al.}(2017)\citenamefont {Veldhorst}, \citenamefont {Eenink}, \citenamefont {Yang},\ and\ \citenamefont {Dzurak}}]{veldhorst2017silicon}%
  \BibitemOpen
  \bibfield  {author} {\bibinfo {author} {\bibfnamefont {M.}~\bibnamefont {Veldhorst}}, \bibinfo {author} {\bibfnamefont {H.~G.}\ \bibnamefont {Eenink}}, \bibinfo {author} {\bibfnamefont {C.-H.}\ \bibnamefont {Yang}},\ and\ \bibinfo {author} {\bibfnamefont {A.~S.}\ \bibnamefont {Dzurak}},\ }\bibfield  {title} {\bibinfo {title} {Silicon cmos architecture for a spin-based quantum computer},\ }\href@noop {} {\bibfield  {journal} {\bibinfo  {journal} {Nature communications}\ }\textbf {\bibinfo {volume} {8}},\ \bibinfo {pages} {1766} (\bibinfo {year} {2017})}\BibitemShut {NoStop}%
\bibitem [{\citenamefont {Boter}\ \emph {et~al.}(2022)\citenamefont {Boter}, \citenamefont {Dehollain}, \citenamefont {Van~Dijk}, \citenamefont {Xu}, \citenamefont {Hensgens}, \citenamefont {Versluis}, \citenamefont {Naus}, \citenamefont {Clarke}, \citenamefont {Veldhorst}, \citenamefont {Sebastiano} \emph {et~al.}}]{boter2022spiderweb}%
  \BibitemOpen
  \bibfield  {author} {\bibinfo {author} {\bibfnamefont {J.~M.}\ \bibnamefont {Boter}}, \bibinfo {author} {\bibfnamefont {J.~P.}\ \bibnamefont {Dehollain}}, \bibinfo {author} {\bibfnamefont {J.~P.}\ \bibnamefont {Van~Dijk}}, \bibinfo {author} {\bibfnamefont {Y.}~\bibnamefont {Xu}}, \bibinfo {author} {\bibfnamefont {T.}~\bibnamefont {Hensgens}}, \bibinfo {author} {\bibfnamefont {R.}~\bibnamefont {Versluis}}, \bibinfo {author} {\bibfnamefont {H.~W.}\ \bibnamefont {Naus}}, \bibinfo {author} {\bibfnamefont {J.~S.}\ \bibnamefont {Clarke}}, \bibinfo {author} {\bibfnamefont {M.}~\bibnamefont {Veldhorst}}, \bibinfo {author} {\bibfnamefont {F.}~\bibnamefont {Sebastiano}}, \emph {et~al.},\ }\bibfield  {title} {\bibinfo {title} {Spiderweb array: A sparse spin-qubit array},\ }\href@noop {} {\bibfield  {journal} {\bibinfo  {journal} {Physical review applied}\ }\textbf {\bibinfo {volume} {18}},\ \bibinfo {pages} {024053} (\bibinfo {year} {2022})}\BibitemShut {NoStop}%
\bibitem [{\citenamefont {Burkard}\ \emph {et~al.}(2023)\citenamefont {Burkard}, \citenamefont {Ladd}, \citenamefont {Pan}, \citenamefont {Nichol},\ and\ \citenamefont {Petta}}]{burkard2023semiconductor}%
  \BibitemOpen
  \bibfield  {author} {\bibinfo {author} {\bibfnamefont {G.}~\bibnamefont {Burkard}}, \bibinfo {author} {\bibfnamefont {T.~D.}\ \bibnamefont {Ladd}}, \bibinfo {author} {\bibfnamefont {A.}~\bibnamefont {Pan}}, \bibinfo {author} {\bibfnamefont {J.~M.}\ \bibnamefont {Nichol}},\ and\ \bibinfo {author} {\bibfnamefont {J.~R.}\ \bibnamefont {Petta}},\ }\bibfield  {title} {\bibinfo {title} {Semiconductor spin qubits},\ }\href@noop {} {\bibfield  {journal} {\bibinfo  {journal} {Reviews of Modern Physics}\ }\textbf {\bibinfo {volume} {95}},\ \bibinfo {pages} {025003} (\bibinfo {year} {2023})}\BibitemShut {NoStop}%
\bibitem [{\citenamefont {Kielpinski}\ \emph {et~al.}(2002)\citenamefont {Kielpinski}, \citenamefont {Monroe},\ and\ \citenamefont {Wineland}}]{Kielpinski2002}%
  \BibitemOpen
  \bibfield  {author} {\bibinfo {author} {\bibfnamefont {D.}~\bibnamefont {Kielpinski}}, \bibinfo {author} {\bibfnamefont {C.}~\bibnamefont {Monroe}},\ and\ \bibinfo {author} {\bibfnamefont {D.~J.}\ \bibnamefont {Wineland}},\ }\bibfield  {title} {\bibinfo {title} {Architecture for a large-scale ion-trap quantum computer},\ }\href {https://doi.org/10.1038/nature00784} {\bibfield  {journal} {\bibinfo  {journal} {Nature}\ }\textbf {\bibinfo {volume} {417}},\ \bibinfo {pages} {709} (\bibinfo {year} {2002})}\BibitemShut {NoStop}%
\bibitem [{\citenamefont {Majer}\ \emph {et~al.}(2007)\citenamefont {Majer}, \citenamefont {Chow}, \citenamefont {Gambetta}, \citenamefont {Koch}, \citenamefont {Johnson}, \citenamefont {Schreier}, \citenamefont {Frunzio}, \citenamefont {Schuster}, \citenamefont {Houck}, \citenamefont {Wallraff}, \citenamefont {Blais}, \citenamefont {Devoret}, \citenamefont {Girvin},\ and\ \citenamefont {Schoelkopf}}]{Majer2007}%
  \BibitemOpen
  \bibfield  {author} {\bibinfo {author} {\bibfnamefont {J.}~\bibnamefont {Majer}}, \bibinfo {author} {\bibfnamefont {J.~M.}\ \bibnamefont {Chow}}, \bibinfo {author} {\bibfnamefont {J.~M.}\ \bibnamefont {Gambetta}}, \bibinfo {author} {\bibfnamefont {J.}~\bibnamefont {Koch}}, \bibinfo {author} {\bibfnamefont {B.~R.}\ \bibnamefont {Johnson}}, \bibinfo {author} {\bibfnamefont {J.~A.}\ \bibnamefont {Schreier}}, \bibinfo {author} {\bibfnamefont {L.}~\bibnamefont {Frunzio}}, \bibinfo {author} {\bibfnamefont {D.~I.}\ \bibnamefont {Schuster}}, \bibinfo {author} {\bibfnamefont {A.~A.}\ \bibnamefont {Houck}}, \bibinfo {author} {\bibfnamefont {A.}~\bibnamefont {Wallraff}}, \bibinfo {author} {\bibfnamefont {A.}~\bibnamefont {Blais}}, \bibinfo {author} {\bibfnamefont {M.~H.}\ \bibnamefont {Devoret}}, \bibinfo {author} {\bibfnamefont {S.~M.}\ \bibnamefont {Girvin}},\ and\ \bibinfo {author} {\bibfnamefont {R.~J.}\ \bibnamefont {Schoelkopf}},\ }\bibfield  {title} {\bibinfo {title} {Coupling superconducting qubits via a
  cavity bus},\ }\href {https://doi.org/10.1038/nature06184} {\bibfield  {journal} {\bibinfo  {journal} {Nature}\ }\textbf {\bibinfo {volume} {449}},\ \bibinfo {pages} {443} (\bibinfo {year} {2007})}\BibitemShut {NoStop}%
\bibitem [{\citenamefont {Bose}(2003)}]{Bose2003}%
  \BibitemOpen
  \bibfield  {author} {\bibinfo {author} {\bibfnamefont {S.}~\bibnamefont {Bose}},\ }\bibfield  {title} {\bibinfo {title} {Quantum communication through an unmodulated spin chain},\ }\href {https://doi.org/10.1103/PhysRevLett.91.207901} {\bibfield  {journal} {\bibinfo  {journal} {Physical Review Letters}\ }\textbf {\bibinfo {volume} {91}},\ \bibinfo {pages} {207901} (\bibinfo {year} {2003})}\BibitemShut {NoStop}%
\bibitem [{\citenamefont {Bose}(2007)}]{Bose2007}%
  \BibitemOpen
  \bibfield  {author} {\bibinfo {author} {\bibfnamefont {S.}~\bibnamefont {Bose}},\ }\bibfield  {title} {\bibinfo {title} {Quantum communication through spin chain dynamics: an introductory overview},\ }\href {https://doi.org/10.1080/00107510701342313} {\bibfield  {journal} {\bibinfo  {journal} {Contemporary Physics}\ }\textbf {\bibinfo {volume} {48}},\ \bibinfo {pages} {13} (\bibinfo {year} {2007})}\BibitemShut {NoStop}%
\bibitem [{\citenamefont {Christandl}\ \emph {et~al.}(2004)\citenamefont {Christandl}, \citenamefont {Datta}, \citenamefont {Ekert},\ and\ \citenamefont {Landahl}}]{Christandl2004}%
  \BibitemOpen
  \bibfield  {author} {\bibinfo {author} {\bibfnamefont {M.}~\bibnamefont {Christandl}}, \bibinfo {author} {\bibfnamefont {N.}~\bibnamefont {Datta}}, \bibinfo {author} {\bibfnamefont {A.}~\bibnamefont {Ekert}},\ and\ \bibinfo {author} {\bibfnamefont {A.~J.}\ \bibnamefont {Landahl}},\ }\bibfield  {title} {\bibinfo {title} {Perfect state transfer in quantum spin networks},\ }\bibfield  {journal} {\bibinfo  {journal} {Physical Review Letters}\ }\textbf {\bibinfo {volume} {92}},\ \href {https://doi.org/10.1103/PhysRevLett.92.187902} {10.1103/PhysRevLett.92.187902} (\bibinfo {year} {2004})\BibitemShut {NoStop}%
\bibitem [{\citenamefont {Christandl}\ \emph {et~al.}(2005)\citenamefont {Christandl}, \citenamefont {Datta}, \citenamefont {Dorlas}, \citenamefont {Ekert}, \citenamefont {Kay},\ and\ \citenamefont {Landahl}}]{Christandl2005}%
  \BibitemOpen
  \bibfield  {author} {\bibinfo {author} {\bibfnamefont {M.}~\bibnamefont {Christandl}}, \bibinfo {author} {\bibfnamefont {N.}~\bibnamefont {Datta}}, \bibinfo {author} {\bibfnamefont {T.~C.}\ \bibnamefont {Dorlas}}, \bibinfo {author} {\bibfnamefont {A.}~\bibnamefont {Ekert}}, \bibinfo {author} {\bibfnamefont {A.}~\bibnamefont {Kay}},\ and\ \bibinfo {author} {\bibfnamefont {A.~J.}\ \bibnamefont {Landahl}},\ }\bibfield  {title} {\bibinfo {title} {Perfect transfer of arbitrary states in quantum spin networks},\ }\bibfield  {journal} {\bibinfo  {journal} {Physical Review A - Atomic, Molecular, and Optical Physics}\ }\textbf {\bibinfo {volume} {71}},\ \href {https://doi.org/10.1103/PhysRevA.71.032312} {10.1103/PhysRevA.71.032312} (\bibinfo {year} {2005})\BibitemShut {NoStop}%
\bibitem [{\citenamefont {Albanese}\ \emph {et~al.}(2004)\citenamefont {Albanese}, \citenamefont {Christandl}, \citenamefont {Datta},\ and\ \citenamefont {Ekert}}]{Albanese2004}%
  \BibitemOpen
  \bibfield  {author} {\bibinfo {author} {\bibfnamefont {C.}~\bibnamefont {Albanese}}, \bibinfo {author} {\bibfnamefont {M.}~\bibnamefont {Christandl}}, \bibinfo {author} {\bibfnamefont {N.}~\bibnamefont {Datta}},\ and\ \bibinfo {author} {\bibfnamefont {A.}~\bibnamefont {Ekert}},\ }\bibfield  {title} {\bibinfo {title} {Mirror inversion of quantum states in linear registers},\ }\bibfield  {journal} {\bibinfo  {journal} {Physical Review Letters}\ }\textbf {\bibinfo {volume} {93}},\ \href {https://doi.org/10.1103/PhysRevLett.93.230502} {10.1103/PhysRevLett.93.230502} (\bibinfo {year} {2004})\BibitemShut {NoStop}%
\bibitem [{\citenamefont {Karbach}\ and\ \citenamefont {Stolze}(2005)}]{Karbach2005}%
  \BibitemOpen
  \bibfield  {author} {\bibinfo {author} {\bibfnamefont {P.}~\bibnamefont {Karbach}}\ and\ \bibinfo {author} {\bibfnamefont {J.}~\bibnamefont {Stolze}},\ }\bibfield  {title} {\bibinfo {title} {Spin chains as perfect quantum state mirrors},\ }\bibfield  {journal} {\bibinfo  {journal} {Physical Review A - Atomic, Molecular, and Optical Physics}\ }\textbf {\bibinfo {volume} {72}},\ \href {https://doi.org/10.1103/PhysRevA.72.030301} {10.1103/PhysRevA.72.030301} (\bibinfo {year} {2005})\BibitemShut {NoStop}%
\bibitem [{\citenamefont {Burgarth}\ and\ \citenamefont {Bose}(2005)}]{burgarth2005conclusive}%
  \BibitemOpen
  \bibfield  {author} {\bibinfo {author} {\bibfnamefont {D.}~\bibnamefont {Burgarth}}\ and\ \bibinfo {author} {\bibfnamefont {S.}~\bibnamefont {Bose}},\ }\bibfield  {title} {\bibinfo {title} {Conclusive and arbitrarily perfect quantum-state transfer using parallel spin-chain channels},\ }\href@noop {} {\bibfield  {journal} {\bibinfo  {journal} {Physical Review A}\ }\textbf {\bibinfo {volume} {71}},\ \bibinfo {pages} {052315} (\bibinfo {year} {2005})}\BibitemShut {NoStop}%
\bibitem [{\citenamefont {Xiang}\ \emph {et~al.}(2024{\natexlab{a}})\citenamefont {Xiang}, \citenamefont {Chen}, \citenamefont {Zhu}, \citenamefont {Song}, \citenamefont {Bao}, \citenamefont {Zhu}, \citenamefont {Jin}, \citenamefont {Wang}, \citenamefont {Xu}, \citenamefont {Zou}, \citenamefont {Li}, \citenamefont {Wang}, \citenamefont {Song}, \citenamefont {Yue}, \citenamefont {Partridge}, \citenamefont {Guo}, \citenamefont {Mondaini}, \citenamefont {Wang},\ and\ \citenamefont {Scalettar}}]{Xiang2024}%
  \BibitemOpen
  \bibfield  {author} {\bibinfo {author} {\bibfnamefont {L.}~\bibnamefont {Xiang}}, \bibinfo {author} {\bibfnamefont {J.}~\bibnamefont {Chen}}, \bibinfo {author} {\bibfnamefont {Z.}~\bibnamefont {Zhu}}, \bibinfo {author} {\bibfnamefont {Z.}~\bibnamefont {Song}}, \bibinfo {author} {\bibfnamefont {Z.}~\bibnamefont {Bao}}, \bibinfo {author} {\bibfnamefont {X.}~\bibnamefont {Zhu}}, \bibinfo {author} {\bibfnamefont {F.}~\bibnamefont {Jin}}, \bibinfo {author} {\bibfnamefont {K.}~\bibnamefont {Wang}}, \bibinfo {author} {\bibfnamefont {S.}~\bibnamefont {Xu}}, \bibinfo {author} {\bibfnamefont {Y.}~\bibnamefont {Zou}}, \bibinfo {author} {\bibfnamefont {H.}~\bibnamefont {Li}}, \bibinfo {author} {\bibfnamefont {Z.}~\bibnamefont {Wang}}, \bibinfo {author} {\bibfnamefont {C.}~\bibnamefont {Song}}, \bibinfo {author} {\bibfnamefont {A.}~\bibnamefont {Yue}}, \bibinfo {author} {\bibfnamefont {J.}~\bibnamefont {Partridge}}, \bibinfo {author} {\bibfnamefont {Q.}~\bibnamefont {Guo}}, \bibinfo {author} {\bibfnamefont
  {R.}~\bibnamefont {Mondaini}}, \bibinfo {author} {\bibfnamefont {H.}~\bibnamefont {Wang}},\ and\ \bibinfo {author} {\bibfnamefont {R.~T.}\ \bibnamefont {Scalettar}},\ }\bibfield  {title} {\bibinfo {title} {Enhanced quantum state transfer by circumventing quantum chaotic behavior},\ }\href {https://doi.org/10.1038/s41467-024-48791-3} {\bibfield  {journal} {\bibinfo  {journal} {Nature Communications}\ }\textbf {\bibinfo {volume} {15}},\ \bibinfo {pages} {4918} (\bibinfo {year} {2024}{\natexlab{a}})}\BibitemShut {NoStop}%
\bibitem [{\citenamefont {Faria}\ \emph {et~al.}(2025)\citenamefont {Faria}, \citenamefont {Nelmes}, \citenamefont {Apollaro}, \citenamefont {Spiller},\ and\ \citenamefont {D’Amico}}]{Faria2025}%
  \BibitemOpen
  \bibfield  {author} {\bibinfo {author} {\bibfnamefont {F.}~\bibnamefont {Faria}}, \bibinfo {author} {\bibfnamefont {C.~C.}\ \bibnamefont {Nelmes}}, \bibinfo {author} {\bibfnamefont {T.~J.~G.}\ \bibnamefont {Apollaro}}, \bibinfo {author} {\bibfnamefont {T.~P.}\ \bibnamefont {Spiller}},\ and\ \bibinfo {author} {\bibfnamefont {I.}~\bibnamefont {D’Amico}},\ }\bibfield  {title} {\bibinfo {title} {Fast and efficient long-distance quantum state transfer in long-range spin- 1 2 models},\ }\href {https://doi.org/10.1088/1367-2630/ae249f} {\bibfield  {journal} {\bibinfo  {journal} {New Journal of Physics}\ }\textbf {\bibinfo {volume} {27}},\ \bibinfo {pages} {124503} (\bibinfo {year} {2025})}\BibitemShut {NoStop}%
\bibitem [{\citenamefont {Bhattacharjya}\ \emph {et~al.}(2024)\citenamefont {Bhattacharjya}, \citenamefont {Monterde},\ and\ \citenamefont {Pal}}]{Bhattacharjya2024}%
  \BibitemOpen
  \bibfield  {author} {\bibinfo {author} {\bibfnamefont {B.}~\bibnamefont {Bhattacharjya}}, \bibinfo {author} {\bibfnamefont {H.}~\bibnamefont {Monterde}},\ and\ \bibinfo {author} {\bibfnamefont {H.}~\bibnamefont {Pal}},\ }\bibfield  {title} {\bibinfo {title} {Quantum walks on blow-up graphs},\ }\bibfield  {journal} {\bibinfo  {journal} {Journal of Physics A: Mathematical and Theoretical}\ }\textbf {\bibinfo {volume} {57}},\ \href {https://doi.org/10.1088/1751-8121/ad6653} {10.1088/1751-8121/ad6653} (\bibinfo {year} {2024})\BibitemShut {NoStop}%
\bibitem [{\citenamefont {Monterde}\ \emph {et~al.}(2025)\citenamefont {Monterde}, \citenamefont {Pal},\ and\ \citenamefont {Kirkland}}]{Monterde2025}%
  \BibitemOpen
  \bibfield  {author} {\bibinfo {author} {\bibfnamefont {H.}~\bibnamefont {Monterde}}, \bibinfo {author} {\bibfnamefont {H.}~\bibnamefont {Pal}},\ and\ \bibinfo {author} {\bibfnamefont {S.}~\bibnamefont {Kirkland}},\ }\bibfield  {title} {\bibinfo {title} {Laplacian quantum walks on blow-up graphs},\ }\bibfield  {journal} {\bibinfo  {journal} {Linear Algebra and Its Applications}\ }\href {https://doi.org/10.1016/j.laa.2025.10.001} {10.1016/j.laa.2025.10.001} (\bibinfo {year} {2025})\BibitemShut {NoStop}%
\bibitem [{\citenamefont {Tashkeev}\ and\ \citenamefont {Zenchuk}(2025)}]{Tashkeev2025}%
  \BibitemOpen
  \bibfield  {author} {\bibinfo {author} {\bibfnamefont {N.~A.}\ \bibnamefont {Tashkeev}}\ and\ \bibinfo {author} {\bibfnamefont {A.~I.}\ \bibnamefont {Zenchuk}},\ }\bibfield  {title} {\bibinfo {title} {Restoring of quantum state transferred along xy-spin chain and entanglement evolution},\ }\bibfield  {journal} {\bibinfo  {journal} {International Journal of Quantum Information}\ }\textbf {\bibinfo {volume} {23}},\ \href {https://doi.org/10.1142/S021974992550011X} {10.1142/S021974992550011X} (\bibinfo {year} {2025})\BibitemShut {NoStop}%
\bibitem [{ban(2011)}]{banchi2011long}%
  \BibitemOpen
  \bibfield  {title} {\bibinfo {title} {Long quantum channels for high-quality entanglement transfer},\ }\href@noop {} {\bibfield  {journal} {\bibinfo  {journal} {New Journal of Physics}\ }\textbf {\bibinfo {volume} {13}},\ \bibinfo {pages} {123006} (\bibinfo {year} {2011})}\BibitemShut {NoStop}%
\bibitem [{\citenamefont {Lewis}\ \emph {et~al.}(2023{\natexlab{a}})\citenamefont {Lewis}, \citenamefont {Moutinho}, \citenamefont {Costa}, \citenamefont {Omar},\ and\ \citenamefont {Bose}}]{Lewis2023}%
  \BibitemOpen
  \bibfield  {author} {\bibinfo {author} {\bibfnamefont {D.}~\bibnamefont {Lewis}}, \bibinfo {author} {\bibfnamefont {J.~P.}\ \bibnamefont {Moutinho}}, \bibinfo {author} {\bibfnamefont {A.~T.}\ \bibnamefont {Costa}}, \bibinfo {author} {\bibfnamefont {Y.}~\bibnamefont {Omar}},\ and\ \bibinfo {author} {\bibfnamefont {S.}~\bibnamefont {Bose}},\ }\bibfield  {title} {\bibinfo {title} {Low-dissipation data bus via coherent quantum dynamics},\ }\href {https://doi.org/10.1103/PhysRevB.108.075405} {\bibfield  {journal} {\bibinfo  {journal} {Physical Review B}\ }\textbf {\bibinfo {volume} {108}},\ \bibinfo {pages} {075405} (\bibinfo {year} {2023}{\natexlab{a}})},\ \bibinfo {note} {for superexchange}\BibitemShut {NoStop}%
\bibitem [{\citenamefont {Wichterich}\ and\ \citenamefont {Bose}(2009)}]{Wichterich2009exploiting}%
  \BibitemOpen
  \bibfield  {author} {\bibinfo {author} {\bibfnamefont {H.}~\bibnamefont {Wichterich}}\ and\ \bibinfo {author} {\bibfnamefont {S.}~\bibnamefont {Bose}},\ }\bibfield  {title} {\bibinfo {title} {Exploiting quench dynamics in spin chains for distant entanglement and quantum communication},\ }\href@noop {} {\bibfield  {journal} {\bibinfo  {journal} {Physical Review A}\ }\textbf {\bibinfo {volume} {79}},\ \bibinfo {pages} {060302(R)} (\bibinfo {year} {2009})}\BibitemShut {NoStop}%
\bibitem [{\citenamefont {Bayat}\ \emph {et~al.}(2010)\citenamefont {Bayat}, \citenamefont {Bose},\ and\ \citenamefont {Sodano}}]{bayat2010entanglement}%
  \BibitemOpen
  \bibfield  {author} {\bibinfo {author} {\bibfnamefont {A.}~\bibnamefont {Bayat}}, \bibinfo {author} {\bibfnamefont {S.}~\bibnamefont {Bose}},\ and\ \bibinfo {author} {\bibfnamefont {P.}~\bibnamefont {Sodano}},\ }\bibfield  {title} {\bibinfo {title} {Entanglement routers using macroscopic singlets},\ }\href@noop {} {\bibfield  {journal} {\bibinfo  {journal} {Physical review letters}\ }\textbf {\bibinfo {volume} {105}},\ \bibinfo {pages} {187204} (\bibinfo {year} {2010})}\BibitemShut {NoStop}%
\bibitem [{\citenamefont {Estarellas}\ \emph {et~al.}(2017)\citenamefont {Estarellas}, \citenamefont {D'Amico},\ and\ \citenamefont {Spiller}}]{estarellas2017robust}%
  \BibitemOpen
  \bibfield  {author} {\bibinfo {author} {\bibfnamefont {M.~P.}\ \bibnamefont {Estarellas}}, \bibinfo {author} {\bibfnamefont {I.}~\bibnamefont {D'Amico}},\ and\ \bibinfo {author} {\bibfnamefont {T.~P.}\ \bibnamefont {Spiller}},\ }\bibfield  {title} {\bibinfo {title} {Robust quantum entanglement generation and generation-plus-storage protocols with spin chains},\ }\href@noop {} {\bibfield  {journal} {\bibinfo  {journal} {Physical Review A}\ }\textbf {\bibinfo {volume} {95}},\ \bibinfo {pages} {042335} (\bibinfo {year} {2017})}\BibitemShut {NoStop}%
\bibitem [{\citenamefont {Divincenzo}\ \emph {et~al.}(1998)\citenamefont {Divincenzo}, \citenamefont {Fuchs}, \citenamefont {Mabuchi}, \citenamefont {Smolin}, \citenamefont {Thapliyal},\ and\ \citenamefont {Uhlmann}}]{Divincenzo1998}%
  \BibitemOpen
  \bibfield  {author} {\bibinfo {author} {\bibfnamefont {D.~P.}\ \bibnamefont {Divincenzo}}, \bibinfo {author} {\bibfnamefont {C.~A.}\ \bibnamefont {Fuchs}}, \bibinfo {author} {\bibfnamefont {H.}~\bibnamefont {Mabuchi}}, \bibinfo {author} {\bibfnamefont {J.~A.}\ \bibnamefont {Smolin}}, \bibinfo {author} {\bibfnamefont {A.}~\bibnamefont {Thapliyal}},\ and\ \bibinfo {author} {\bibfnamefont {A.}~\bibnamefont {Uhlmann}},\ }\href@noop {} {\emph {\bibinfo {title} {Entanglement of Assistance}}},\ \bibinfo {type} {Tech. Rep.}\ (\bibinfo {year} {1998})\BibitemShut {NoStop}%
\bibitem [{\citenamefont {Verstraete}\ \emph {et~al.}(2004)\citenamefont {Verstraete}, \citenamefont {Popp},\ and\ \citenamefont {Cirac}}]{Verstraete2004}%
  \BibitemOpen
  \bibfield  {author} {\bibinfo {author} {\bibfnamefont {F.}~\bibnamefont {Verstraete}}, \bibinfo {author} {\bibfnamefont {M.}~\bibnamefont {Popp}},\ and\ \bibinfo {author} {\bibfnamefont {J.~I.}\ \bibnamefont {Cirac}},\ }\bibfield  {title} {\bibinfo {title} {Entanglement versus correlations in spin systems},\ }\href {https://doi.org/10.1103/PhysRevLett.92.027901} {\bibfield  {journal} {\bibinfo  {journal} {Physical Review Letters}\ }\textbf {\bibinfo {volume} {92}},\ \bibinfo {pages} {027901} (\bibinfo {year} {2004})}\BibitemShut {NoStop}%
\bibitem [{\citenamefont {Vaucher}\ \emph {et~al.}(2005)\citenamefont {Vaucher}, \citenamefont {Burgarth},\ and\ \citenamefont {Bose}}]{vaucher2005arbitrarily}%
  \BibitemOpen
  \bibfield  {author} {\bibinfo {author} {\bibfnamefont {B.}~\bibnamefont {Vaucher}}, \bibinfo {author} {\bibfnamefont {D.}~\bibnamefont {Burgarth}},\ and\ \bibinfo {author} {\bibfnamefont {S.}~\bibnamefont {Bose}},\ }\bibfield  {title} {\bibinfo {title} {Arbitrarily perfect quantum communication using unmodulated spin chains, a collaborative approach},\ }\href@noop {} {\bibfield  {journal} {\bibinfo  {journal} {Journal of Optics B: Quantum and Semiclassical Optics}\ }\textbf {\bibinfo {volume} {7}},\ \bibinfo {pages} {S356} (\bibinfo {year} {2005})}\BibitemShut {NoStop}%
\bibitem [{\citenamefont {Shizume}\ \emph {et~al.}(2007)\citenamefont {Shizume}, \citenamefont {Jacobs}, \citenamefont {Burgarth},\ and\ \citenamefont {Bose}}]{shizume2007quantum}%
  \BibitemOpen
  \bibfield  {author} {\bibinfo {author} {\bibfnamefont {K.}~\bibnamefont {Shizume}}, \bibinfo {author} {\bibfnamefont {K.}~\bibnamefont {Jacobs}}, \bibinfo {author} {\bibfnamefont {D.}~\bibnamefont {Burgarth}},\ and\ \bibinfo {author} {\bibfnamefont {S.}~\bibnamefont {Bose}},\ }\bibfield  {title} {\bibinfo {title} {Quantum communication via a continuously monitored dual spin chain},\ }\href@noop {} {\bibfield  {journal} {\bibinfo  {journal} {Physical Review A}\ }\textbf {\bibinfo {volume} {75}},\ \bibinfo {pages} {062328} (\bibinfo {year} {2007})}\BibitemShut {NoStop}%
\bibitem [{\citenamefont {Fel'dman}\ \emph {et~al.}(2026)\citenamefont {Fel'dman}, \citenamefont {Wu},\ and\ \citenamefont {Zenchuk}}]{Feldman2026_PPST}%
  \BibitemOpen
  \bibfield  {author} {\bibinfo {author} {\bibfnamefont {E.~B.}\ \bibnamefont {Fel'dman}}, \bibinfo {author} {\bibfnamefont {J.}~\bibnamefont {Wu}},\ and\ \bibinfo {author} {\bibfnamefont {A.~I.}\ \bibnamefont {Zenchuk}},\ }\href {https://arxiv.org/abs/2509.10100} {\bibinfo {title} {Perfect pure quantum state transfer via state restoring and ancilla measurement}} (\bibinfo {year} {2026}),\ \Eprint {https://arxiv.org/abs/2509.10100} {arXiv:2509.10100 [quant-ph]} \BibitemShut {NoStop}%
\bibitem [{\citenamefont {Fel'dman}\ and\ \citenamefont {Zenchuk}(2026)}]{Feldman2026_MQST}%
  \BibitemOpen
  \bibfield  {author} {\bibinfo {author} {\bibfnamefont {E.~B.}\ \bibnamefont {Fel'dman}}\ and\ \bibinfo {author} {\bibfnamefont {A.~I.}\ \bibnamefont {Zenchuk}},\ }\href {https://arxiv.org/abs/2605.13051} {\bibinfo {title} {Measurement-based quantum state transfer and restoring via spin-1/2 chain interacting with environment}} (\bibinfo {year} {2026}),\ \Eprint {https://arxiv.org/abs/2605.13051} {arXiv:2605.13051 [quant-ph]} \BibitemShut {NoStop}%
\bibitem [{\citenamefont {Xiang}\ \emph {et~al.}(2024{\natexlab{b}})\citenamefont {Xiang}, \citenamefont {Chen}, \citenamefont {Zhu}, \citenamefont {Song}, \citenamefont {Bao}, \citenamefont {Zhu}, \citenamefont {Jin}, \citenamefont {Wang}, \citenamefont {Xu}, \citenamefont {Zou} \emph {et~al.}}]{xiang2024enhanced}%
  \BibitemOpen
  \bibfield  {author} {\bibinfo {author} {\bibfnamefont {L.}~\bibnamefont {Xiang}}, \bibinfo {author} {\bibfnamefont {J.}~\bibnamefont {Chen}}, \bibinfo {author} {\bibfnamefont {Z.}~\bibnamefont {Zhu}}, \bibinfo {author} {\bibfnamefont {Z.}~\bibnamefont {Song}}, \bibinfo {author} {\bibfnamefont {Z.}~\bibnamefont {Bao}}, \bibinfo {author} {\bibfnamefont {X.}~\bibnamefont {Zhu}}, \bibinfo {author} {\bibfnamefont {F.}~\bibnamefont {Jin}}, \bibinfo {author} {\bibfnamefont {K.}~\bibnamefont {Wang}}, \bibinfo {author} {\bibfnamefont {S.}~\bibnamefont {Xu}}, \bibinfo {author} {\bibfnamefont {Y.}~\bibnamefont {Zou}}, \emph {et~al.},\ }\bibfield  {title} {\bibinfo {title} {Enhanced quantum state transfer by circumventing quantum chaotic behavior},\ }\href@noop {} {\bibfield  {journal} {\bibinfo  {journal} {Nature Communications}\ }\textbf {\bibinfo {volume} {15}},\ \bibinfo {pages} {4918} (\bibinfo {year} {2024}{\natexlab{b}})}\BibitemShut {NoStop}%
\bibitem [{\citenamefont {Li}\ \emph {et~al.}(2018)\citenamefont {Li}, \citenamefont {Ma}, \citenamefont {Han}, \citenamefont {Chen}, \citenamefont {Xu}, \citenamefont {Cai}, \citenamefont {Wang}, \citenamefont {Song}, \citenamefont {Xue}, \citenamefont {Yin} \emph {et~al.}}]{li2018perfect}%
  \BibitemOpen
  \bibfield  {author} {\bibinfo {author} {\bibfnamefont {X.}~\bibnamefont {Li}}, \bibinfo {author} {\bibfnamefont {Y.}~\bibnamefont {Ma}}, \bibinfo {author} {\bibfnamefont {J.}~\bibnamefont {Han}}, \bibinfo {author} {\bibfnamefont {T.}~\bibnamefont {Chen}}, \bibinfo {author} {\bibfnamefont {Y.}~\bibnamefont {Xu}}, \bibinfo {author} {\bibfnamefont {W.}~\bibnamefont {Cai}}, \bibinfo {author} {\bibfnamefont {H.}~\bibnamefont {Wang}}, \bibinfo {author} {\bibfnamefont {Y.-P.}\ \bibnamefont {Song}}, \bibinfo {author} {\bibfnamefont {Z.-Y.}\ \bibnamefont {Xue}}, \bibinfo {author} {\bibfnamefont {Z.-q.}\ \bibnamefont {Yin}}, \emph {et~al.},\ }\bibfield  {title} {\bibinfo {title} {Perfect quantum state transfer in a superconducting qubit chain with parametrically tunable couplings},\ }\href@noop {} {\bibfield  {journal} {\bibinfo  {journal} {Physical Review Applied}\ }\textbf {\bibinfo {volume} {10}},\ \bibinfo {pages} {054009} (\bibinfo {year} {2018})}\BibitemShut {NoStop}%
\bibitem [{\citenamefont {Kandel}\ \emph {et~al.}(2019)\citenamefont {Kandel}, \citenamefont {Qiao}, \citenamefont {Fallahi}, \citenamefont {Gardner}, \citenamefont {Manfra},\ and\ \citenamefont {Nichol}}]{kandel2019coherent}%
  \BibitemOpen
  \bibfield  {author} {\bibinfo {author} {\bibfnamefont {Y.~P.}\ \bibnamefont {Kandel}}, \bibinfo {author} {\bibfnamefont {H.}~\bibnamefont {Qiao}}, \bibinfo {author} {\bibfnamefont {S.}~\bibnamefont {Fallahi}}, \bibinfo {author} {\bibfnamefont {G.~C.}\ \bibnamefont {Gardner}}, \bibinfo {author} {\bibfnamefont {M.~J.}\ \bibnamefont {Manfra}},\ and\ \bibinfo {author} {\bibfnamefont {J.~M.}\ \bibnamefont {Nichol}},\ }\bibfield  {title} {\bibinfo {title} {Coherent spin-state transfer via heisenberg exchange},\ }\href@noop {} {\bibfield  {journal} {\bibinfo  {journal} {Nature}\ }\textbf {\bibinfo {volume} {573}},\ \bibinfo {pages} {553} (\bibinfo {year} {2019})}\BibitemShut {NoStop}%
\bibitem [{\citenamefont {Nakajima}\ \emph {et~al.}(2018)\citenamefont {Nakajima}, \citenamefont {Delbecq}, \citenamefont {Otsuka}, \citenamefont {Amaha}, \citenamefont {Yoneda}, \citenamefont {Noiri}, \citenamefont {Takeda}, \citenamefont {Allison}, \citenamefont {Ludwig}, \citenamefont {Wieck} \emph {et~al.}}]{nakajima2018coherent}%
  \BibitemOpen
  \bibfield  {author} {\bibinfo {author} {\bibfnamefont {T.}~\bibnamefont {Nakajima}}, \bibinfo {author} {\bibfnamefont {M.~R.}\ \bibnamefont {Delbecq}}, \bibinfo {author} {\bibfnamefont {T.}~\bibnamefont {Otsuka}}, \bibinfo {author} {\bibfnamefont {S.}~\bibnamefont {Amaha}}, \bibinfo {author} {\bibfnamefont {J.}~\bibnamefont {Yoneda}}, \bibinfo {author} {\bibfnamefont {A.}~\bibnamefont {Noiri}}, \bibinfo {author} {\bibfnamefont {K.}~\bibnamefont {Takeda}}, \bibinfo {author} {\bibfnamefont {G.}~\bibnamefont {Allison}}, \bibinfo {author} {\bibfnamefont {A.}~\bibnamefont {Ludwig}}, \bibinfo {author} {\bibfnamefont {A.~D.}\ \bibnamefont {Wieck}}, \emph {et~al.},\ }\bibfield  {title} {\bibinfo {title} {Coherent transfer of electron spin correlations assisted by dephasing noise},\ }\href@noop {} {\bibfield  {journal} {\bibinfo  {journal} {Nature communications}\ }\textbf {\bibinfo {volume} {9}},\ \bibinfo {pages} {2133} (\bibinfo {year} {2018})}\BibitemShut {NoStop}%
\bibitem [{\citenamefont {Sigillito}\ \emph {et~al.}(2019)\citenamefont {Sigillito}, \citenamefont {Gullans}, \citenamefont {Edge}, \citenamefont {Borselli},\ and\ \citenamefont {Petta}}]{sigillito2019coherent}%
  \BibitemOpen
  \bibfield  {author} {\bibinfo {author} {\bibfnamefont {A.}~\bibnamefont {Sigillito}}, \bibinfo {author} {\bibfnamefont {M.}~\bibnamefont {Gullans}}, \bibinfo {author} {\bibfnamefont {L.}~\bibnamefont {Edge}}, \bibinfo {author} {\bibfnamefont {M.}~\bibnamefont {Borselli}},\ and\ \bibinfo {author} {\bibfnamefont {J.}~\bibnamefont {Petta}},\ }\bibfield  {title} {\bibinfo {title} {Coherent transfer of quantum information in a silicon double quantum dot using resonant swap gates},\ }\href@noop {} {\bibfield  {journal} {\bibinfo  {journal} {npj Quantum Information}\ }\textbf {\bibinfo {volume} {5}},\ \bibinfo {pages} {110} (\bibinfo {year} {2019})}\BibitemShut {NoStop}%
\bibitem [{\citenamefont {Fari{\~n}a}\ \emph {et~al.}(2025)\citenamefont {Fari{\~n}a}, \citenamefont {Jirovec}, \citenamefont {Zhang}, \citenamefont {Morozova}, \citenamefont {Oosterhout}, \citenamefont {Reale}, \citenamefont {Hsiao}, \citenamefont {Scappucci}, \citenamefont {Veldhorst},\ and\ \citenamefont {Vandersypen}}]{farina2025site}%
  \BibitemOpen
  \bibfield  {author} {\bibinfo {author} {\bibfnamefont {P.~C.}\ \bibnamefont {Fari{\~n}a}}, \bibinfo {author} {\bibfnamefont {D.}~\bibnamefont {Jirovec}}, \bibinfo {author} {\bibfnamefont {X.}~\bibnamefont {Zhang}}, \bibinfo {author} {\bibfnamefont {E.}~\bibnamefont {Morozova}}, \bibinfo {author} {\bibfnamefont {S.~D.}\ \bibnamefont {Oosterhout}}, \bibinfo {author} {\bibfnamefont {S.}~\bibnamefont {Reale}}, \bibinfo {author} {\bibfnamefont {T.-K.}\ \bibnamefont {Hsiao}}, \bibinfo {author} {\bibfnamefont {G.}~\bibnamefont {Scappucci}}, \bibinfo {author} {\bibfnamefont {M.}~\bibnamefont {Veldhorst}},\ and\ \bibinfo {author} {\bibfnamefont {L.~M.}\ \bibnamefont {Vandersypen}},\ }\bibfield  {title} {\bibinfo {title} {Site-resolved magnon and triplon dynamics on a programmable quantum dot spin ladder},\ }\href@noop {} {\bibfield  {journal} {\bibinfo  {journal} {arXiv preprint arXiv:2506.08663}\ } (\bibinfo {year} {2025})}\BibitemShut {NoStop}%
\bibitem [{\citenamefont {Soriano}\ \emph {et~al.}(2026)\citenamefont {Soriano}, \citenamefont {Manoj~Kumar}, \citenamefont {Gerbaud}, \citenamefont {Savoyant}, \citenamefont {Dassonneville}, \citenamefont {Vezin}, \citenamefont {Jeannin}, \citenamefont {Orio}, \citenamefont {Fourmigu{\'e}},\ and\ \citenamefont {Bertaina}}]{soriano2026exploring}%
  \BibitemOpen
  \bibfield  {author} {\bibinfo {author} {\bibfnamefont {L.}~\bibnamefont {Soriano}}, \bibinfo {author} {\bibfnamefont {A.}~\bibnamefont {Manoj~Kumar}}, \bibinfo {author} {\bibfnamefont {G.}~\bibnamefont {Gerbaud}}, \bibinfo {author} {\bibfnamefont {A.}~\bibnamefont {Savoyant}}, \bibinfo {author} {\bibfnamefont {R.}~\bibnamefont {Dassonneville}}, \bibinfo {author} {\bibfnamefont {H.}~\bibnamefont {Vezin}}, \bibinfo {author} {\bibfnamefont {O.}~\bibnamefont {Jeannin}}, \bibinfo {author} {\bibfnamefont {M.}~\bibnamefont {Orio}}, \bibinfo {author} {\bibfnamefont {M.}~\bibnamefont {Fourmigu{\'e}}},\ and\ \bibinfo {author} {\bibfnamefont {S.}~\bibnamefont {Bertaina}},\ }\bibfield  {title} {\bibinfo {title} {Exploring electron spin dynamics in spin chains using defects as a quantum probe},\ }\href@noop {} {\bibfield  {journal} {\bibinfo  {journal} {Nature Communications}\ }\textbf {\bibinfo {volume} {17}},\ \bibinfo {pages} {4046} (\bibinfo {year} {2026})}\BibitemShut {NoStop}%
\bibitem [{\citenamefont {Coden}\ \emph {et~al.}(2025)\citenamefont {Coden}, \citenamefont {Osenda},\ and\ \citenamefont {Ferrón}}]{AcostaCoden2025}%
  \BibitemOpen
  \bibfield  {author} {\bibinfo {author} {\bibfnamefont {D.~S.~A.}\ \bibnamefont {Coden}}, \bibinfo {author} {\bibfnamefont {O.}~\bibnamefont {Osenda}},\ and\ \bibinfo {author} {\bibfnamefont {A.}~\bibnamefont {Ferrón}},\ }\bibfield  {title} {\bibinfo {title} {Quantum state transfer in a magnetic atoms chain using a scanning tunneling microscope},\ }\bibfield  {journal} {\bibinfo  {journal} {Journal of Physics B: Atomic, Molecular and Optical Physics}\ }\textbf {\bibinfo {volume} {58}},\ \href {https://doi.org/10.1088/1361-6455/ad9a30} {10.1088/1361-6455/ad9a30} (\bibinfo {year} {2025})\BibitemShut {NoStop}%
\bibitem [{\citenamefont {Yao}\ \emph {et~al.}(2011)\citenamefont {Yao}, \citenamefont {Jiang}, \citenamefont {Gorshkov}, \citenamefont {Gong}, \citenamefont {Zhai}, \citenamefont {Duan},\ and\ \citenamefont {Lukin}}]{yao2011robust}%
  \BibitemOpen
  \bibfield  {author} {\bibinfo {author} {\bibfnamefont {N.~Y.}\ \bibnamefont {Yao}}, \bibinfo {author} {\bibfnamefont {L.}~\bibnamefont {Jiang}}, \bibinfo {author} {\bibfnamefont {A.~V.}\ \bibnamefont {Gorshkov}}, \bibinfo {author} {\bibfnamefont {Z.-X.}\ \bibnamefont {Gong}}, \bibinfo {author} {\bibfnamefont {A.}~\bibnamefont {Zhai}}, \bibinfo {author} {\bibfnamefont {L.-M.}\ \bibnamefont {Duan}},\ and\ \bibinfo {author} {\bibfnamefont {M.~D.}\ \bibnamefont {Lukin}},\ }\bibfield  {title} {\bibinfo {title} {Robust quantum state transfer in random unpolarized spin chains},\ }\href@noop {} {\bibfield  {journal} {\bibinfo  {journal} {Physical Review Letters}\ }\textbf {\bibinfo {volume} {106}},\ \bibinfo {pages} {040505} (\bibinfo {year} {2011})}\BibitemShut {NoStop}%
\bibitem [{\citenamefont {Duan}\ \emph {et~al.}(2003)\citenamefont {Duan}, \citenamefont {Demler},\ and\ \citenamefont {Lukin}}]{duan2003controlling}%
  \BibitemOpen
  \bibfield  {author} {\bibinfo {author} {\bibfnamefont {L.-M.}\ \bibnamefont {Duan}}, \bibinfo {author} {\bibfnamefont {E.}~\bibnamefont {Demler}},\ and\ \bibinfo {author} {\bibfnamefont {M.~D.}\ \bibnamefont {Lukin}},\ }\bibfield  {title} {\bibinfo {title} {Controlling spin exchange interactions of ultracold atoms in optical lattices},\ }\href@noop {} {\bibfield  {journal} {\bibinfo  {journal} {Physical review letters}\ }\textbf {\bibinfo {volume} {91}},\ \bibinfo {pages} {090402} (\bibinfo {year} {2003})}\BibitemShut {NoStop}%
\bibitem [{\citenamefont {Banchi}\ \emph {et~al.}(2011)\citenamefont {Banchi}, \citenamefont {Bayat}, \citenamefont {Verrucchi},\ and\ \citenamefont {Bose}}]{banchi2011nonperturbative}%
  \BibitemOpen
  \bibfield  {author} {\bibinfo {author} {\bibfnamefont {L.}~\bibnamefont {Banchi}}, \bibinfo {author} {\bibfnamefont {A.}~\bibnamefont {Bayat}}, \bibinfo {author} {\bibfnamefont {P.}~\bibnamefont {Verrucchi}},\ and\ \bibinfo {author} {\bibfnamefont {S.}~\bibnamefont {Bose}},\ }\bibfield  {title} {\bibinfo {title} {Nonperturbative entangling gates between distant qubits using uniform cold atom chains},\ }\href@noop {} {\bibfield  {journal} {\bibinfo  {journal} {Physical review letters}\ }\textbf {\bibinfo {volume} {106}},\ \bibinfo {pages} {140501} (\bibinfo {year} {2011})}\BibitemShut {NoStop}%
\bibitem [{jur(2014)}]{jurcevic2014quasiparticle}%
  \BibitemOpen
  \bibfield  {title} {\bibinfo {title} {Quasiparticle engineering and entanglement propagation in a quantum many-body system},\ }\href@noop {} {\bibfield  {journal} {\bibinfo  {journal} {Nature}\ }\textbf {\bibinfo {volume} {511}},\ \bibinfo {pages} {202} (\bibinfo {year} {2014})}\BibitemShut {NoStop}%
\bibitem [{\citenamefont {Richerme}\ \emph {et~al.}(2014)\citenamefont {Richerme}, \citenamefont {Gong}, \citenamefont {Lee}, \citenamefont {Senko}, \citenamefont {Smith}, \citenamefont {Foss-Feig}, \citenamefont {Michalakis}, \citenamefont {Gorshkov},\ and\ \citenamefont {Monroe}}]{richerme2014non}%
  \BibitemOpen
  \bibfield  {author} {\bibinfo {author} {\bibfnamefont {P.}~\bibnamefont {Richerme}}, \bibinfo {author} {\bibfnamefont {Z.-X.}\ \bibnamefont {Gong}}, \bibinfo {author} {\bibfnamefont {A.}~\bibnamefont {Lee}}, \bibinfo {author} {\bibfnamefont {C.}~\bibnamefont {Senko}}, \bibinfo {author} {\bibfnamefont {J.}~\bibnamefont {Smith}}, \bibinfo {author} {\bibfnamefont {M.}~\bibnamefont {Foss-Feig}}, \bibinfo {author} {\bibfnamefont {S.}~\bibnamefont {Michalakis}}, \bibinfo {author} {\bibfnamefont {A.~V.}\ \bibnamefont {Gorshkov}},\ and\ \bibinfo {author} {\bibfnamefont {C.}~\bibnamefont {Monroe}},\ }\bibfield  {title} {\bibinfo {title} {Non-local propagation of correlations in quantum systems with long-range interactions},\ }\href@noop {} {\bibfield  {journal} {\bibinfo  {journal} {Nature}\ }\textbf {\bibinfo {volume} {511}},\ \bibinfo {pages} {198} (\bibinfo {year} {2014})}\BibitemShut {NoStop}%
\bibitem [{\citenamefont {Sarkar}\ \emph {et~al.}(2026{\natexlab{a}})\citenamefont {Sarkar}, \citenamefont {Pointon},\ and\ \citenamefont {Bose}}]{sarkar2026spin}%
  \BibitemOpen
  \bibfield  {author} {\bibinfo {author} {\bibfnamefont {M.}~\bibnamefont {Sarkar}}, \bibinfo {author} {\bibfnamefont {T.}~\bibnamefont {Pointon}},\ and\ \bibinfo {author} {\bibfnamefont {S.}~\bibnamefont {Bose}},\ }\bibfield  {title} {\bibinfo {title} {Spin chain quantum communication on a trapped-ion processor},\ }\href@noop {} {\bibfield  {journal} {\bibinfo  {journal} {arXiv preprint arXiv:2607.12999}\ } (\bibinfo {year} {2026}{\natexlab{a}})}\BibitemShut {NoStop}%
\bibitem [{\citenamefont {Sgroi}\ \emph {et~al.}(2025)\citenamefont {Sgroi}, \citenamefont {Zicari}, \citenamefont {Imparato},\ and\ \citenamefont {Paternostro}}]{sgroi2025reinforcement}%
  \BibitemOpen
  \bibfield  {author} {\bibinfo {author} {\bibfnamefont {S.}~\bibnamefont {Sgroi}}, \bibinfo {author} {\bibfnamefont {G.}~\bibnamefont {Zicari}}, \bibinfo {author} {\bibfnamefont {A.}~\bibnamefont {Imparato}},\ and\ \bibinfo {author} {\bibfnamefont {M.}~\bibnamefont {Paternostro}},\ }\bibfield  {title} {\bibinfo {title} {A reinforcement learning approach to the design of quantum chains for optimal energy and state transfer},\ }\href@noop {} {\bibfield  {journal} {\bibinfo  {journal} {Machine Learning: Science and Technology}\ }\textbf {\bibinfo {volume} {6}},\ \bibinfo {pages} {015012} (\bibinfo {year} {2025})}\BibitemShut {NoStop}%
\bibitem [{\citenamefont {Michel}\ \emph {et~al.}(2026)\citenamefont {Michel}, \citenamefont {Werner},\ and\ \citenamefont {Riera}}]{michel2026quantum}%
  \BibitemOpen
  \bibfield  {author} {\bibinfo {author} {\bibfnamefont {O.}~\bibnamefont {Michel}}, \bibinfo {author} {\bibfnamefont {M.}~\bibnamefont {Werner}},\ and\ \bibinfo {author} {\bibfnamefont {A.}~\bibnamefont {Riera}},\ }\bibfield  {title} {\bibinfo {title} {A quantum state transfer protocol with ising hamiltonians},\ }\href@noop {} {\bibfield  {journal} {\bibinfo  {journal} {Quantum Science and Technology}\ }\textbf {\bibinfo {volume} {11}},\ \bibinfo {pages} {035006} (\bibinfo {year} {2026})}\BibitemShut {NoStop}%
\bibitem [{\citenamefont {Sarkar}\ \emph {et~al.}(2026{\natexlab{b}})\citenamefont {Sarkar}, \citenamefont {Ghosh}, \citenamefont {Smith}, \citenamefont {Myronov},\ and\ \citenamefont {Bose}}]{sarkar2026all}%
  \BibitemOpen
  \bibfield  {author} {\bibinfo {author} {\bibfnamefont {M.}~\bibnamefont {Sarkar}}, \bibinfo {author} {\bibfnamefont {R.}~\bibnamefont {Ghosh}}, \bibinfo {author} {\bibfnamefont {C.~G.}\ \bibnamefont {Smith}}, \bibinfo {author} {\bibfnamefont {M.}~\bibnamefont {Myronov}},\ and\ \bibinfo {author} {\bibfnamefont {S.}~\bibnamefont {Bose}},\ }\bibfield  {title} {\bibinfo {title} {All-electric quantum state transfer via spin-orbit phase matching},\ }\href@noop {} {\bibfield  {journal} {\bibinfo  {journal} {arXiv preprint arXiv:2605.13976}\ } (\bibinfo {year} {2026}{\natexlab{b}})}\BibitemShut {NoStop}%
\bibitem [{\citenamefont {Avellino}\ \emph {et~al.}(2006)\citenamefont {Avellino}, \citenamefont {Fisher},\ and\ \citenamefont {Bose}}]{avellino2006quantum}%
  \BibitemOpen
  \bibfield  {author} {\bibinfo {author} {\bibfnamefont {M.}~\bibnamefont {Avellino}}, \bibinfo {author} {\bibfnamefont {A.}~\bibnamefont {Fisher}},\ and\ \bibinfo {author} {\bibfnamefont {S.}~\bibnamefont {Bose}},\ }\bibfield  {title} {\bibinfo {title} {Quantum communication in spin systems with long-range interactions},\ }\href@noop {} {\bibfield  {journal} {\bibinfo  {journal} {Physical Review A—Atomic, Molecular, and Optical Physics}\ }\textbf {\bibinfo {volume} {74}},\ \bibinfo {pages} {012321} (\bibinfo {year} {2006})}\BibitemShut {NoStop}%
\bibitem [{\citenamefont {Lewis}\ \emph {et~al.}(2023{\natexlab{b}})\citenamefont {Lewis}, \citenamefont {Banchi}, \citenamefont {Hong~Teoh}, \citenamefont {Islam},\ and\ \citenamefont {Bose}}]{lewis2023ion}%
  \BibitemOpen
  \bibfield  {author} {\bibinfo {author} {\bibfnamefont {D.}~\bibnamefont {Lewis}}, \bibinfo {author} {\bibfnamefont {L.}~\bibnamefont {Banchi}}, \bibinfo {author} {\bibfnamefont {Y.}~\bibnamefont {Hong~Teoh}}, \bibinfo {author} {\bibfnamefont {R.}~\bibnamefont {Islam}},\ and\ \bibinfo {author} {\bibfnamefont {S.}~\bibnamefont {Bose}},\ }\bibfield  {title} {\bibinfo {title} {Ion trap long-range xy model for quantum state transfer and optimal spatial search},\ }\href@noop {} {\bibfield  {journal} {\bibinfo  {journal} {Quantum Science and Technology}\ }\textbf {\bibinfo {volume} {8}},\ \bibinfo {pages} {035025} (\bibinfo {year} {2023}{\natexlab{b}})}\BibitemShut {NoStop}%
\bibitem [{\citenamefont {Ahuja}\ \emph {et~al.}(2026{\natexlab{a}})\citenamefont {Ahuja}, \citenamefont {Konar}, \citenamefont {Lakkaraju},\ and\ \citenamefont {Sen}}]{ahuja2026expediting}%
  \BibitemOpen
  \bibfield  {author} {\bibinfo {author} {\bibfnamefont {S.}~\bibnamefont {Ahuja}}, \bibinfo {author} {\bibfnamefont {T.~K.}\ \bibnamefont {Konar}}, \bibinfo {author} {\bibfnamefont {L.~G.~C.}\ \bibnamefont {Lakkaraju}},\ and\ \bibinfo {author} {\bibfnamefont {A.}~\bibnamefont {Sen}},\ }\bibfield  {title} {\bibinfo {title} {Expediting quantum state transfer through the long-range extended xy model},\ }\href@noop {} {\bibfield  {journal} {\bibinfo  {journal} {Physical Review A}\ }\textbf {\bibinfo {volume} {113}},\ \bibinfo {pages} {022426} (\bibinfo {year} {2026}{\natexlab{a}})}\BibitemShut {NoStop}%
\bibitem [{\citenamefont {Marshall}\ \emph {et~al.}(2026)\citenamefont {Marshall}, \citenamefont {Nelmes}, \citenamefont {Apollaro}, \citenamefont {Spiller},\ and\ \citenamefont {D'Amico}}]{marshall2026distinct}%
  \BibitemOpen
  \bibfield  {author} {\bibinfo {author} {\bibfnamefont {E.}~\bibnamefont {Marshall}}, \bibinfo {author} {\bibfnamefont {C.}~\bibnamefont {Nelmes}}, \bibinfo {author} {\bibfnamefont {T.}~\bibnamefont {Apollaro}}, \bibinfo {author} {\bibfnamefont {T.}~\bibnamefont {Spiller}},\ and\ \bibinfo {author} {\bibfnamefont {I.}~\bibnamefont {D'Amico}},\ }\bibfield  {title} {\bibinfo {title} {Distinct modes of quantum information transfer in power-law long-range spin networks},\ }\href@noop {} {\bibfield  {journal} {\bibinfo  {journal} {arXiv preprint arXiv:2608.19057}\ } (\bibinfo {year} {2026})}\BibitemShut {NoStop}%
\bibitem [{\citenamefont {Hamonic}\ \emph {et~al.}(2025)\citenamefont {Hamonic}, \citenamefont {Nurizzo}, \citenamefont {Nath}, \citenamefont {Dartiailh}, \citenamefont {Elhomsy}, \citenamefont {Fragnol}, \citenamefont {Martinez}, \citenamefont {Julliard}, \citenamefont {Cardoso~Paz}, \citenamefont {Ouvrier-Buffet} \emph {et~al.}}]{hamonic2025combining}%
  \BibitemOpen
  \bibfield  {author} {\bibinfo {author} {\bibfnamefont {P.}~\bibnamefont {Hamonic}}, \bibinfo {author} {\bibfnamefont {M.}~\bibnamefont {Nurizzo}}, \bibinfo {author} {\bibfnamefont {J.}~\bibnamefont {Nath}}, \bibinfo {author} {\bibfnamefont {M.~C.}\ \bibnamefont {Dartiailh}}, \bibinfo {author} {\bibfnamefont {V.}~\bibnamefont {Elhomsy}}, \bibinfo {author} {\bibfnamefont {M.}~\bibnamefont {Fragnol}}, \bibinfo {author} {\bibfnamefont {B.}~\bibnamefont {Martinez}}, \bibinfo {author} {\bibfnamefont {P.-L.}\ \bibnamefont {Julliard}}, \bibinfo {author} {\bibfnamefont {B.}~\bibnamefont {Cardoso~Paz}}, \bibinfo {author} {\bibfnamefont {M.}~\bibnamefont {Ouvrier-Buffet}}, \emph {et~al.},\ }\bibfield  {title} {\bibinfo {title} {Combining multiplexed gate-based readout and isolated cmos quantum dot arrays},\ }\href@noop {} {\bibfield  {journal} {\bibinfo  {journal} {Nature Communications}\ }\textbf {\bibinfo {volume} {16}},\ \bibinfo {pages} {6323} (\bibinfo {year} {2025})}\BibitemShut {NoStop}%
\bibitem [{\citenamefont {Bakr}\ \emph {et~al.}(2025)\citenamefont {Bakr}, \citenamefont {Fasciati}, \citenamefont {Cao}, \citenamefont {Campanaro}, \citenamefont {Wills}, \citenamefont {Alghadeer}, \citenamefont {Piscitelli}, \citenamefont {Shteynas}, \citenamefont {Chidambaram},\ and\ \citenamefont {Leek}}]{bakr2025multiplexed}%
  \BibitemOpen
  \bibfield  {author} {\bibinfo {author} {\bibfnamefont {M.}~\bibnamefont {Bakr}}, \bibinfo {author} {\bibfnamefont {S.~D.}\ \bibnamefont {Fasciati}}, \bibinfo {author} {\bibfnamefont {S.}~\bibnamefont {Cao}}, \bibinfo {author} {\bibfnamefont {G.}~\bibnamefont {Campanaro}}, \bibinfo {author} {\bibfnamefont {J.}~\bibnamefont {Wills}}, \bibinfo {author} {\bibfnamefont {M.}~\bibnamefont {Alghadeer}}, \bibinfo {author} {\bibfnamefont {M.}~\bibnamefont {Piscitelli}}, \bibinfo {author} {\bibfnamefont {B.}~\bibnamefont {Shteynas}}, \bibinfo {author} {\bibfnamefont {V.}~\bibnamefont {Chidambaram}},\ and\ \bibinfo {author} {\bibfnamefont {P.~J.}\ \bibnamefont {Leek}},\ }\bibfield  {title} {\bibinfo {title} {Multiplexed readout of superconducting qubits using a three-dimensional reentrant-cavity filter},\ }\href@noop {} {\bibfield  {journal} {\bibinfo  {journal} {Physical Review Applied}\ }\textbf {\bibinfo {volume} {23}},\ \bibinfo {pages} {054089} (\bibinfo {year} {2025})}\BibitemShut {NoStop}%
\bibitem [{\citenamefont {Deist}\ \emph {et~al.}(2022)\citenamefont {Deist}, \citenamefont {Lu}, \citenamefont {Ho}, \citenamefont {Pasha}, \citenamefont {Zeiher}, \citenamefont {Yan},\ and\ \citenamefont {Stamper-Kurn}}]{deist2022mid}%
  \BibitemOpen
  \bibfield  {author} {\bibinfo {author} {\bibfnamefont {E.}~\bibnamefont {Deist}}, \bibinfo {author} {\bibfnamefont {Y.-H.}\ \bibnamefont {Lu}}, \bibinfo {author} {\bibfnamefont {J.}~\bibnamefont {Ho}}, \bibinfo {author} {\bibfnamefont {M.~K.}\ \bibnamefont {Pasha}}, \bibinfo {author} {\bibfnamefont {J.}~\bibnamefont {Zeiher}}, \bibinfo {author} {\bibfnamefont {Z.}~\bibnamefont {Yan}},\ and\ \bibinfo {author} {\bibfnamefont {D.~M.}\ \bibnamefont {Stamper-Kurn}},\ }\bibfield  {title} {\bibinfo {title} {Mid-circuit cavity measurement in a neutral atom array},\ }\href@noop {} {\bibfield  {journal} {\bibinfo  {journal} {Physical Review Letters}\ }\textbf {\bibinfo {volume} {129}},\ \bibinfo {pages} {203602} (\bibinfo {year} {2022})}\BibitemShut {NoStop}%
\bibitem [{\citenamefont {Bluvstein}\ \emph {et~al.}(2026)\citenamefont {Bluvstein}, \citenamefont {Geim}, \citenamefont {Li}, \citenamefont {Evered}, \citenamefont {Bonilla~Ataides}, \citenamefont {Baranes}, \citenamefont {Gu}, \citenamefont {Manovitz}, \citenamefont {Xu}, \citenamefont {Kalinowski} \emph {et~al.}}]{bluvstein2026fault}%
  \BibitemOpen
  \bibfield  {author} {\bibinfo {author} {\bibfnamefont {D.}~\bibnamefont {Bluvstein}}, \bibinfo {author} {\bibfnamefont {A.~A.}\ \bibnamefont {Geim}}, \bibinfo {author} {\bibfnamefont {S.~H.}\ \bibnamefont {Li}}, \bibinfo {author} {\bibfnamefont {S.~J.}\ \bibnamefont {Evered}}, \bibinfo {author} {\bibfnamefont {J.~P.}\ \bibnamefont {Bonilla~Ataides}}, \bibinfo {author} {\bibfnamefont {G.}~\bibnamefont {Baranes}}, \bibinfo {author} {\bibfnamefont {A.}~\bibnamefont {Gu}}, \bibinfo {author} {\bibfnamefont {T.}~\bibnamefont {Manovitz}}, \bibinfo {author} {\bibfnamefont {M.}~\bibnamefont {Xu}}, \bibinfo {author} {\bibfnamefont {M.}~\bibnamefont {Kalinowski}}, \emph {et~al.},\ }\bibfield  {title} {\bibinfo {title} {A fault-tolerant neutral-atom architecture for universal quantum computation},\ }\href@noop {} {\bibfield  {journal} {\bibinfo  {journal} {Nature}\ }\textbf {\bibinfo {volume} {649}},\ \bibinfo {pages} {39} (\bibinfo {year} {2026})}\BibitemShut {NoStop}%
\bibitem [{\citenamefont {Motlakunta}\ \emph {et~al.}(2024)\citenamefont {Motlakunta}, \citenamefont {Kotibhaskar}, \citenamefont {Shih}, \citenamefont {Vogliano}, \citenamefont {McLaren}, \citenamefont {Hahn}, \citenamefont {Zhu}, \citenamefont {Habl{\"u}tzel},\ and\ \citenamefont {Islam}}]{motlakunta2024preserving}%
  \BibitemOpen
  \bibfield  {author} {\bibinfo {author} {\bibfnamefont {S.}~\bibnamefont {Motlakunta}}, \bibinfo {author} {\bibfnamefont {N.}~\bibnamefont {Kotibhaskar}}, \bibinfo {author} {\bibfnamefont {C.-Y.}\ \bibnamefont {Shih}}, \bibinfo {author} {\bibfnamefont {A.}~\bibnamefont {Vogliano}}, \bibinfo {author} {\bibfnamefont {D.}~\bibnamefont {McLaren}}, \bibinfo {author} {\bibfnamefont {L.}~\bibnamefont {Hahn}}, \bibinfo {author} {\bibfnamefont {J.}~\bibnamefont {Zhu}}, \bibinfo {author} {\bibfnamefont {R.}~\bibnamefont {Habl{\"u}tzel}},\ and\ \bibinfo {author} {\bibfnamefont {R.}~\bibnamefont {Islam}},\ }\bibfield  {title} {\bibinfo {title} {Preserving a qubit during state-destroying operations on an adjacent qubit at a few micrometers distance},\ }\href@noop {} {\bibfield  {journal} {\bibinfo  {journal} {Nature Communications}\ }\textbf {\bibinfo {volume} {15}},\ \bibinfo {pages} {6575} (\bibinfo {year} {2024})}\BibitemShut {NoStop}%
\bibitem [{\citenamefont {Fisher}\ \emph {et~al.}(2023)\citenamefont {Fisher}, \citenamefont {Khemani}, \citenamefont {Nahum},\ and\ \citenamefont {Vijay}}]{fisher2023random}%
  \BibitemOpen
  \bibfield  {author} {\bibinfo {author} {\bibfnamefont {M.~P.}\ \bibnamefont {Fisher}}, \bibinfo {author} {\bibfnamefont {V.}~\bibnamefont {Khemani}}, \bibinfo {author} {\bibfnamefont {A.}~\bibnamefont {Nahum}},\ and\ \bibinfo {author} {\bibfnamefont {S.}~\bibnamefont {Vijay}},\ }\bibfield  {title} {\bibinfo {title} {Random quantum circuits},\ }\href@noop {} {\bibfield  {journal} {\bibinfo  {journal} {Annual Review of Condensed Matter Physics}\ }\textbf {\bibinfo {volume} {14}},\ \bibinfo {pages} {335} (\bibinfo {year} {2023})}\BibitemShut {NoStop}%
\bibitem [{\citenamefont {Potter}\ and\ \citenamefont {Vasseur}(2022)}]{potter2022entanglement}%
  \BibitemOpen
  \bibfield  {author} {\bibinfo {author} {\bibfnamefont {A.~C.}\ \bibnamefont {Potter}}\ and\ \bibinfo {author} {\bibfnamefont {R.}~\bibnamefont {Vasseur}},\ }\bibfield  {title} {\bibinfo {title} {Entanglement dynamics in hybrid quantum circuits},\ }in\ \href@noop {} {\emph {\bibinfo {booktitle} {Entanglement in Spin Chains: From Theory to Quantum Technology Applications}}}\ (\bibinfo  {publisher} {Springer},\ \bibinfo {year} {2022})\ pp.\ \bibinfo {pages} {211--249}\BibitemShut {NoStop}%
\bibitem [{\citenamefont {Lunt}\ \emph {et~al.}(2022)\citenamefont {Lunt}, \citenamefont {Richter},\ and\ \citenamefont {Pal}}]{lunt2022quantum}%
  \BibitemOpen
  \bibfield  {author} {\bibinfo {author} {\bibfnamefont {O.}~\bibnamefont {Lunt}}, \bibinfo {author} {\bibfnamefont {J.}~\bibnamefont {Richter}},\ and\ \bibinfo {author} {\bibfnamefont {A.}~\bibnamefont {Pal}},\ }\bibfield  {title} {\bibinfo {title} {Quantum simulation using noisy unitary circuits and measurements},\ }in\ \href@noop {} {\emph {\bibinfo {booktitle} {Entanglement in Spin Chains: From Theory to Quantum Technology Applications}}}\ (\bibinfo  {publisher} {Springer},\ \bibinfo {year} {2022})\ pp.\ \bibinfo {pages} {251--284}\BibitemShut {NoStop}%
\bibitem [{\citenamefont {B{\"a}umer}\ \emph {et~al.}(2024)\citenamefont {B{\"a}umer}, \citenamefont {Tripathi}, \citenamefont {Wang}, \citenamefont {Rall}, \citenamefont {Chen}, \citenamefont {Majumder}, \citenamefont {Seif},\ and\ \citenamefont {Minev}}]{baumer2024efficient}%
  \BibitemOpen
  \bibfield  {author} {\bibinfo {author} {\bibfnamefont {E.}~\bibnamefont {B{\"a}umer}}, \bibinfo {author} {\bibfnamefont {V.}~\bibnamefont {Tripathi}}, \bibinfo {author} {\bibfnamefont {D.~S.}\ \bibnamefont {Wang}}, \bibinfo {author} {\bibfnamefont {P.}~\bibnamefont {Rall}}, \bibinfo {author} {\bibfnamefont {E.~H.}\ \bibnamefont {Chen}}, \bibinfo {author} {\bibfnamefont {S.}~\bibnamefont {Majumder}}, \bibinfo {author} {\bibfnamefont {A.}~\bibnamefont {Seif}},\ and\ \bibinfo {author} {\bibfnamefont {Z.~K.}\ \bibnamefont {Minev}},\ }\bibfield  {title} {\bibinfo {title} {Efficient long-range entanglement using dynamic circuits},\ }\href@noop {} {\bibfield  {journal} {\bibinfo  {journal} {PRX quantum}\ }\textbf {\bibinfo {volume} {5}},\ \bibinfo {pages} {030339} (\bibinfo {year} {2024})}\BibitemShut {NoStop}%
\bibitem [{\citenamefont {Sang}\ \emph {et~al.}(2023)\citenamefont {Sang}, \citenamefont {Li}, \citenamefont {Hsieh},\ and\ \citenamefont {Yoshida}}]{sang2023ultrafast}%
  \BibitemOpen
  \bibfield  {author} {\bibinfo {author} {\bibfnamefont {S.}~\bibnamefont {Sang}}, \bibinfo {author} {\bibfnamefont {Z.}~\bibnamefont {Li}}, \bibinfo {author} {\bibfnamefont {T.~H.}\ \bibnamefont {Hsieh}},\ and\ \bibinfo {author} {\bibfnamefont {B.}~\bibnamefont {Yoshida}},\ }\bibfield  {title} {\bibinfo {title} {Ultrafast entanglement dynamics in monitored quantum circuits},\ }\href@noop {} {\bibfield  {journal} {\bibinfo  {journal} {PRX Quantum}\ }\textbf {\bibinfo {volume} {4}},\ \bibinfo {pages} {040332} (\bibinfo {year} {2023})}\BibitemShut {NoStop}%
\bibitem [{\citenamefont {Ahuja}\ \emph {et~al.}(2026{\natexlab{b}})\citenamefont {Ahuja}, \citenamefont {Agarwal},\ and\ \citenamefont {De}}]{ahuja2026dynamically}%
  \BibitemOpen
  \bibfield  {author} {\bibinfo {author} {\bibfnamefont {S.}~\bibnamefont {Ahuja}}, \bibinfo {author} {\bibfnamefont {K.~D.}\ \bibnamefont {Agarwal}},\ and\ \bibinfo {author} {\bibfnamefont {A.~S.}\ \bibnamefont {De}},\ }\bibfield  {title} {\bibinfo {title} {Dynamically frozen long-distance entanglement via non-hermitian pt-symmetric systems},\ }\href@noop {} {\bibfield  {journal} {\bibinfo  {journal} {arXiv preprint arXiv:2606.14177}\ } (\bibinfo {year} {2026}{\natexlab{b}})}\BibitemShut {NoStop}%
\bibitem [{\citenamefont {Bao}\ \emph {et~al.}(2024)\citenamefont {Bao}, \citenamefont {Block},\ and\ \citenamefont {Altman}}]{bao2024finite}%
  \BibitemOpen
  \bibfield  {author} {\bibinfo {author} {\bibfnamefont {Y.}~\bibnamefont {Bao}}, \bibinfo {author} {\bibfnamefont {M.}~\bibnamefont {Block}},\ and\ \bibinfo {author} {\bibfnamefont {E.}~\bibnamefont {Altman}},\ }\bibfield  {title} {\bibinfo {title} {Finite-time teleportation phase transition in random quantum circuits},\ }\href@noop {} {\bibfield  {journal} {\bibinfo  {journal} {Physical Review Letters}\ }\textbf {\bibinfo {volume} {132}},\ \bibinfo {pages} {030401} (\bibinfo {year} {2024})}\BibitemShut {NoStop}%
\bibitem [{\citenamefont {Hoke}\ \emph {et~al.}(2023)\citenamefont {Hoke}, \citenamefont {Ippoliti}, \citenamefont {Rosenberg}, \citenamefont {Abanin}, \citenamefont {Acharya}, \citenamefont {Andersen}, \citenamefont {Ansmann}, \citenamefont {Arute}, \citenamefont {Arya}, \citenamefont {Asfaw}, \citenamefont {Atalaya}, \citenamefont {Bardin}, \citenamefont {Bengtsson}, \citenamefont {Bortoli}, \citenamefont {Bourassa}, \citenamefont {Bovaird}, \citenamefont {Brill}, \citenamefont {Broughton}, \citenamefont {Buckley}, \citenamefont {Buell}, \citenamefont {Burger}, \citenamefont {Burkett}, \citenamefont {Bushnell}, \citenamefont {Chen}, \citenamefont {Chiaro}, \citenamefont {Chik}, \citenamefont {Cogan}, \citenamefont {Collins}, \citenamefont {Conner}, \citenamefont {Courtney}, \citenamefont {Crook}, \citenamefont {Curtin}, \citenamefont {Dau}, \citenamefont {Debroy}, \citenamefont {Barba}, \citenamefont {Demura}, \citenamefont {Paolo}, \citenamefont {Drozdov}, \citenamefont {Dunsworth}, \citenamefont {Eppens},
  \citenamefont {Erickson}, \citenamefont {Farhi}, \citenamefont {Fatemi}, \citenamefont {Ferreira}, \citenamefont {Burgos}, \citenamefont {Forati}, \citenamefont {Fowler}, \citenamefont {Foxen}, \citenamefont {Giang}, \citenamefont {Gidney}, \citenamefont {Gilboa}, \citenamefont {Giustina}, \citenamefont {Gosula}, \citenamefont {Gross}, \citenamefont {Habegger}, \citenamefont {Hamilton}, \citenamefont {Hansen}, \citenamefont {Harrigan}, \citenamefont {Harrington}, \citenamefont {Heu}, \citenamefont {Hoffmann}, \citenamefont {Hong}, \citenamefont {Huang}, \citenamefont {Huff}, \citenamefont {Huggins}, \citenamefont {Isakov}, \citenamefont {Iveland}, \citenamefont {Jeffrey}, \citenamefont {Jiang}, \citenamefont {Jones}, \citenamefont {Juhas}, \citenamefont {Kafri}, \citenamefont {Kechedzhi}, \citenamefont {Khattar}, \citenamefont {Khezri}, \citenamefont {Kieferová}, \citenamefont {Kim}, \citenamefont {Kitaev}, \citenamefont {Klimov}, \citenamefont {Klots}, \citenamefont {Korotkov}, \citenamefont {Kostritsa},
  \citenamefont {Kreikebaum}, \citenamefont {Landhuis}, \citenamefont {Laptev}, \citenamefont {Lau}, \citenamefont {Laws}, \citenamefont {Lee}, \citenamefont {Lee}, \citenamefont {Lensky}, \citenamefont {Lester}, \citenamefont {Lill}, \citenamefont {Liu}, \citenamefont {Locharla}, \citenamefont {Martin}, \citenamefont {McClean}, \citenamefont {McEwen}, \citenamefont {Miao}, \citenamefont {Mieszala}, \citenamefont {Montazeri}, \citenamefont {Morvan}, \citenamefont {Movassagh}, \citenamefont {Mruczkiewicz}, \citenamefont {Neeley}, \citenamefont {Neill}, \citenamefont {Nersisyan}, \citenamefont {Newman}, \citenamefont {Ng}, \citenamefont {Nguyen}, \citenamefont {Nguyen}, \citenamefont {Niu}, \citenamefont {O’Brien}, \citenamefont {Omonije}, \citenamefont {Opremcak}, \citenamefont {Petukhov}, \citenamefont {Potter}, \citenamefont {Pryadko}, \citenamefont {Quintana}, \citenamefont {Rocque}, \citenamefont {Rubin}, \citenamefont {Saei}, \citenamefont {Sank}, \citenamefont {Sankaragomathi}, \citenamefont
  {Satzinger}, \citenamefont {Schurkus}, \citenamefont {Schuster}, \citenamefont {Shearn}, \citenamefont {Shorter}, \citenamefont {Shutty}, \citenamefont {Shvarts}, \citenamefont {Skruzny}, \citenamefont {Smith}, \citenamefont {Somma}, \citenamefont {Sterling}, \citenamefont {Strain}, \citenamefont {Szalay}, \citenamefont {Torres}, \citenamefont {Vidal}, \citenamefont {Villalonga}, \citenamefont {Heidweiller}, \citenamefont {White}, \citenamefont {Woo}, \citenamefont {Xing}, \citenamefont {Yao}, \citenamefont {Yeh}, \citenamefont {Yoo}, \citenamefont {Young}, \citenamefont {Zalcman}, \citenamefont {Zhang}, \citenamefont {Zhu}, \citenamefont {Zobrist}, \citenamefont {Neven}, \citenamefont {Babbush}, \citenamefont {Bacon}, \citenamefont {Boixo}, \citenamefont {Hilton}, \citenamefont {Lucero}, \citenamefont {Megrant}, \citenamefont {Kelly}, \citenamefont {Chen}, \citenamefont {Smelyanskiy}, \citenamefont {Mi}, \citenamefont {Khemani},\ and\ \citenamefont {Roushan}}]{Hoke2023}%
  \BibitemOpen
  \bibfield  {author} {\bibinfo {author} {\bibfnamefont {J.~C.}\ \bibnamefont {Hoke}}, \bibinfo {author} {\bibfnamefont {M.}~\bibnamefont {Ippoliti}}, \bibinfo {author} {\bibfnamefont {E.}~\bibnamefont {Rosenberg}}, \bibinfo {author} {\bibfnamefont {D.}~\bibnamefont {Abanin}}, \bibinfo {author} {\bibfnamefont {R.}~\bibnamefont {Acharya}}, \bibinfo {author} {\bibfnamefont {T.~I.}\ \bibnamefont {Andersen}}, \bibinfo {author} {\bibfnamefont {M.}~\bibnamefont {Ansmann}}, \bibinfo {author} {\bibfnamefont {F.}~\bibnamefont {Arute}}, \bibinfo {author} {\bibfnamefont {K.}~\bibnamefont {Arya}}, \bibinfo {author} {\bibfnamefont {A.}~\bibnamefont {Asfaw}}, \bibinfo {author} {\bibfnamefont {J.}~\bibnamefont {Atalaya}}, \bibinfo {author} {\bibfnamefont {J.~C.}\ \bibnamefont {Bardin}}, \bibinfo {author} {\bibfnamefont {A.}~\bibnamefont {Bengtsson}}, \bibinfo {author} {\bibfnamefont {G.}~\bibnamefont {Bortoli}}, \bibinfo {author} {\bibfnamefont {A.}~\bibnamefont {Bourassa}}, \bibinfo {author} {\bibfnamefont
  {J.}~\bibnamefont {Bovaird}}, \bibinfo {author} {\bibfnamefont {L.}~\bibnamefont {Brill}}, \bibinfo {author} {\bibfnamefont {M.}~\bibnamefont {Broughton}}, \bibinfo {author} {\bibfnamefont {B.~B.}\ \bibnamefont {Buckley}}, \bibinfo {author} {\bibfnamefont {D.~A.}\ \bibnamefont {Buell}}, \bibinfo {author} {\bibfnamefont {T.}~\bibnamefont {Burger}}, \bibinfo {author} {\bibfnamefont {B.}~\bibnamefont {Burkett}}, \bibinfo {author} {\bibfnamefont {N.}~\bibnamefont {Bushnell}}, \bibinfo {author} {\bibfnamefont {Z.}~\bibnamefont {Chen}}, \bibinfo {author} {\bibfnamefont {B.}~\bibnamefont {Chiaro}}, \bibinfo {author} {\bibfnamefont {D.}~\bibnamefont {Chik}}, \bibinfo {author} {\bibfnamefont {J.}~\bibnamefont {Cogan}}, \bibinfo {author} {\bibfnamefont {R.}~\bibnamefont {Collins}}, \bibinfo {author} {\bibfnamefont {P.}~\bibnamefont {Conner}}, \bibinfo {author} {\bibfnamefont {W.}~\bibnamefont {Courtney}}, \bibinfo {author} {\bibfnamefont {A.~L.}\ \bibnamefont {Crook}}, \bibinfo {author} {\bibfnamefont
  {B.}~\bibnamefont {Curtin}}, \bibinfo {author} {\bibfnamefont {A.~G.}\ \bibnamefont {Dau}}, \bibinfo {author} {\bibfnamefont {D.~M.}\ \bibnamefont {Debroy}}, \bibinfo {author} {\bibfnamefont {A.~D.~T.}\ \bibnamefont {Barba}}, \bibinfo {author} {\bibfnamefont {S.}~\bibnamefont {Demura}}, \bibinfo {author} {\bibfnamefont {A.~D.}\ \bibnamefont {Paolo}}, \bibinfo {author} {\bibfnamefont {I.~K.}\ \bibnamefont {Drozdov}}, \bibinfo {author} {\bibfnamefont {A.}~\bibnamefont {Dunsworth}}, \bibinfo {author} {\bibfnamefont {D.}~\bibnamefont {Eppens}}, \bibinfo {author} {\bibfnamefont {C.}~\bibnamefont {Erickson}}, \bibinfo {author} {\bibfnamefont {E.}~\bibnamefont {Farhi}}, \bibinfo {author} {\bibfnamefont {R.}~\bibnamefont {Fatemi}}, \bibinfo {author} {\bibfnamefont {V.~S.}\ \bibnamefont {Ferreira}}, \bibinfo {author} {\bibfnamefont {L.~F.}\ \bibnamefont {Burgos}}, \bibinfo {author} {\bibfnamefont {E.}~\bibnamefont {Forati}}, \bibinfo {author} {\bibfnamefont {A.~G.}\ \bibnamefont {Fowler}}, \bibinfo {author}
  {\bibfnamefont {B.}~\bibnamefont {Foxen}}, \bibinfo {author} {\bibfnamefont {W.}~\bibnamefont {Giang}}, \bibinfo {author} {\bibfnamefont {C.}~\bibnamefont {Gidney}}, \bibinfo {author} {\bibfnamefont {D.}~\bibnamefont {Gilboa}}, \bibinfo {author} {\bibfnamefont {M.}~\bibnamefont {Giustina}}, \bibinfo {author} {\bibfnamefont {R.}~\bibnamefont {Gosula}}, \bibinfo {author} {\bibfnamefont {J.~A.}\ \bibnamefont {Gross}}, \bibinfo {author} {\bibfnamefont {S.}~\bibnamefont {Habegger}}, \bibinfo {author} {\bibfnamefont {M.~C.}\ \bibnamefont {Hamilton}}, \bibinfo {author} {\bibfnamefont {M.}~\bibnamefont {Hansen}}, \bibinfo {author} {\bibfnamefont {M.~P.}\ \bibnamefont {Harrigan}}, \bibinfo {author} {\bibfnamefont {S.~D.}\ \bibnamefont {Harrington}}, \bibinfo {author} {\bibfnamefont {P.}~\bibnamefont {Heu}}, \bibinfo {author} {\bibfnamefont {M.~R.}\ \bibnamefont {Hoffmann}}, \bibinfo {author} {\bibfnamefont {S.}~\bibnamefont {Hong}}, \bibinfo {author} {\bibfnamefont {T.}~\bibnamefont {Huang}}, \bibinfo {author}
  {\bibfnamefont {A.}~\bibnamefont {Huff}}, \bibinfo {author} {\bibfnamefont {W.~J.}\ \bibnamefont {Huggins}}, \bibinfo {author} {\bibfnamefont {S.~V.}\ \bibnamefont {Isakov}}, \bibinfo {author} {\bibfnamefont {J.}~\bibnamefont {Iveland}}, \bibinfo {author} {\bibfnamefont {E.}~\bibnamefont {Jeffrey}}, \bibinfo {author} {\bibfnamefont {Z.}~\bibnamefont {Jiang}}, \bibinfo {author} {\bibfnamefont {C.}~\bibnamefont {Jones}}, \bibinfo {author} {\bibfnamefont {P.}~\bibnamefont {Juhas}}, \bibinfo {author} {\bibfnamefont {D.}~\bibnamefont {Kafri}}, \bibinfo {author} {\bibfnamefont {K.}~\bibnamefont {Kechedzhi}}, \bibinfo {author} {\bibfnamefont {T.}~\bibnamefont {Khattar}}, \bibinfo {author} {\bibfnamefont {M.}~\bibnamefont {Khezri}}, \bibinfo {author} {\bibfnamefont {M.}~\bibnamefont {Kieferová}}, \bibinfo {author} {\bibfnamefont {S.}~\bibnamefont {Kim}}, \bibinfo {author} {\bibfnamefont {A.}~\bibnamefont {Kitaev}}, \bibinfo {author} {\bibfnamefont {P.~V.}\ \bibnamefont {Klimov}}, \bibinfo {author} {\bibfnamefont
  {A.~R.}\ \bibnamefont {Klots}}, \bibinfo {author} {\bibfnamefont {A.~N.}\ \bibnamefont {Korotkov}}, \bibinfo {author} {\bibfnamefont {F.}~\bibnamefont {Kostritsa}}, \bibinfo {author} {\bibfnamefont {J.~M.}\ \bibnamefont {Kreikebaum}}, \bibinfo {author} {\bibfnamefont {D.}~\bibnamefont {Landhuis}}, \bibinfo {author} {\bibfnamefont {P.}~\bibnamefont {Laptev}}, \bibinfo {author} {\bibfnamefont {K.-M.}\ \bibnamefont {Lau}}, \bibinfo {author} {\bibfnamefont {L.}~\bibnamefont {Laws}}, \bibinfo {author} {\bibfnamefont {J.}~\bibnamefont {Lee}}, \bibinfo {author} {\bibfnamefont {K.~W.}\ \bibnamefont {Lee}}, \bibinfo {author} {\bibfnamefont {Y.~D.}\ \bibnamefont {Lensky}}, \bibinfo {author} {\bibfnamefont {B.~J.}\ \bibnamefont {Lester}}, \bibinfo {author} {\bibfnamefont {A.~T.}\ \bibnamefont {Lill}}, \bibinfo {author} {\bibfnamefont {W.}~\bibnamefont {Liu}}, \bibinfo {author} {\bibfnamefont {A.}~\bibnamefont {Locharla}}, \bibinfo {author} {\bibfnamefont {O.}~\bibnamefont {Martin}}, \bibinfo {author} {\bibfnamefont
  {J.~R.}\ \bibnamefont {McClean}}, \bibinfo {author} {\bibfnamefont {M.}~\bibnamefont {McEwen}}, \bibinfo {author} {\bibfnamefont {K.~C.}\ \bibnamefont {Miao}}, \bibinfo {author} {\bibfnamefont {A.}~\bibnamefont {Mieszala}}, \bibinfo {author} {\bibfnamefont {S.}~\bibnamefont {Montazeri}}, \bibinfo {author} {\bibfnamefont {A.}~\bibnamefont {Morvan}}, \bibinfo {author} {\bibfnamefont {R.}~\bibnamefont {Movassagh}}, \bibinfo {author} {\bibfnamefont {W.}~\bibnamefont {Mruczkiewicz}}, \bibinfo {author} {\bibfnamefont {M.}~\bibnamefont {Neeley}}, \bibinfo {author} {\bibfnamefont {C.}~\bibnamefont {Neill}}, \bibinfo {author} {\bibfnamefont {A.}~\bibnamefont {Nersisyan}}, \bibinfo {author} {\bibfnamefont {M.}~\bibnamefont {Newman}}, \bibinfo {author} {\bibfnamefont {J.~H.}\ \bibnamefont {Ng}}, \bibinfo {author} {\bibfnamefont {A.}~\bibnamefont {Nguyen}}, \bibinfo {author} {\bibfnamefont {M.}~\bibnamefont {Nguyen}}, \bibinfo {author} {\bibfnamefont {M.~Y.}\ \bibnamefont {Niu}}, \bibinfo {author} {\bibfnamefont
  {T.~E.}\ \bibnamefont {O’Brien}}, \bibinfo {author} {\bibfnamefont {S.}~\bibnamefont {Omonije}}, \bibinfo {author} {\bibfnamefont {A.}~\bibnamefont {Opremcak}}, \bibinfo {author} {\bibfnamefont {A.}~\bibnamefont {Petukhov}}, \bibinfo {author} {\bibfnamefont {R.}~\bibnamefont {Potter}}, \bibinfo {author} {\bibfnamefont {L.~P.}\ \bibnamefont {Pryadko}}, \bibinfo {author} {\bibfnamefont {C.}~\bibnamefont {Quintana}}, \bibinfo {author} {\bibfnamefont {C.}~\bibnamefont {Rocque}}, \bibinfo {author} {\bibfnamefont {N.~C.}\ \bibnamefont {Rubin}}, \bibinfo {author} {\bibfnamefont {N.}~\bibnamefont {Saei}}, \bibinfo {author} {\bibfnamefont {D.}~\bibnamefont {Sank}}, \bibinfo {author} {\bibfnamefont {K.}~\bibnamefont {Sankaragomathi}}, \bibinfo {author} {\bibfnamefont {K.~J.}\ \bibnamefont {Satzinger}}, \bibinfo {author} {\bibfnamefont {H.~F.}\ \bibnamefont {Schurkus}}, \bibinfo {author} {\bibfnamefont {C.}~\bibnamefont {Schuster}}, \bibinfo {author} {\bibfnamefont {M.~J.}\ \bibnamefont {Shearn}}, \bibinfo {author}
  {\bibfnamefont {A.}~\bibnamefont {Shorter}}, \bibinfo {author} {\bibfnamefont {N.}~\bibnamefont {Shutty}}, \bibinfo {author} {\bibfnamefont {V.}~\bibnamefont {Shvarts}}, \bibinfo {author} {\bibfnamefont {J.}~\bibnamefont {Skruzny}}, \bibinfo {author} {\bibfnamefont {W.~C.}\ \bibnamefont {Smith}}, \bibinfo {author} {\bibfnamefont {R.}~\bibnamefont {Somma}}, \bibinfo {author} {\bibfnamefont {G.}~\bibnamefont {Sterling}}, \bibinfo {author} {\bibfnamefont {D.}~\bibnamefont {Strain}}, \bibinfo {author} {\bibfnamefont {M.}~\bibnamefont {Szalay}}, \bibinfo {author} {\bibfnamefont {A.}~\bibnamefont {Torres}}, \bibinfo {author} {\bibfnamefont {G.}~\bibnamefont {Vidal}}, \bibinfo {author} {\bibfnamefont {B.}~\bibnamefont {Villalonga}}, \bibinfo {author} {\bibfnamefont {C.~V.}\ \bibnamefont {Heidweiller}}, \bibinfo {author} {\bibfnamefont {T.}~\bibnamefont {White}}, \bibinfo {author} {\bibfnamefont {B.~W.~K.}\ \bibnamefont {Woo}}, \bibinfo {author} {\bibfnamefont {C.}~\bibnamefont {Xing}}, \bibinfo {author}
  {\bibfnamefont {Z.~J.}\ \bibnamefont {Yao}}, \bibinfo {author} {\bibfnamefont {P.}~\bibnamefont {Yeh}}, \bibinfo {author} {\bibfnamefont {J.}~\bibnamefont {Yoo}}, \bibinfo {author} {\bibfnamefont {G.}~\bibnamefont {Young}}, \bibinfo {author} {\bibfnamefont {A.}~\bibnamefont {Zalcman}}, \bibinfo {author} {\bibfnamefont {Y.}~\bibnamefont {Zhang}}, \bibinfo {author} {\bibfnamefont {N.}~\bibnamefont {Zhu}}, \bibinfo {author} {\bibfnamefont {N.}~\bibnamefont {Zobrist}}, \bibinfo {author} {\bibfnamefont {H.}~\bibnamefont {Neven}}, \bibinfo {author} {\bibfnamefont {R.}~\bibnamefont {Babbush}}, \bibinfo {author} {\bibfnamefont {D.}~\bibnamefont {Bacon}}, \bibinfo {author} {\bibfnamefont {S.}~\bibnamefont {Boixo}}, \bibinfo {author} {\bibfnamefont {J.}~\bibnamefont {Hilton}}, \bibinfo {author} {\bibfnamefont {E.}~\bibnamefont {Lucero}}, \bibinfo {author} {\bibfnamefont {A.}~\bibnamefont {Megrant}}, \bibinfo {author} {\bibfnamefont {J.}~\bibnamefont {Kelly}}, \bibinfo {author} {\bibfnamefont {Y.}~\bibnamefont
  {Chen}}, \bibinfo {author} {\bibfnamefont {V.}~\bibnamefont {Smelyanskiy}}, \bibinfo {author} {\bibfnamefont {X.}~\bibnamefont {Mi}}, \bibinfo {author} {\bibfnamefont {V.}~\bibnamefont {Khemani}},\ and\ \bibinfo {author} {\bibfnamefont {P.}~\bibnamefont {Roushan}},\ }\bibfield  {title} {\bibinfo {title} {Measurement-induced entanglement and teleportation on a noisy quantum processor},\ }\href {https://doi.org/10.1038/s41586-023-06505-7} {\bibfield  {journal} {\bibinfo  {journal} {Nature}\ }\textbf {\bibinfo {volume} {622}},\ \bibinfo {pages} {481} (\bibinfo {year} {2023})}\BibitemShut {NoStop}%
\bibitem [{\citenamefont {Lo}\ and\ \citenamefont {Popescu}(2001)}]{Lo2001}%
  \BibitemOpen
  \bibfield  {author} {\bibinfo {author} {\bibfnamefont {H.~K.}\ \bibnamefont {Lo}}\ and\ \bibinfo {author} {\bibfnamefont {S.}~\bibnamefont {Popescu}},\ }\bibfield  {title} {\bibinfo {title} {Concentrating entanglement by local actions: Beyond mean values},\ }\href {https://doi.org/10.1103/PhysRevA.63.022301} {\bibfield  {journal} {\bibinfo  {journal} {Physical Review A - Atomic, Molecular, and Optical Physics}\ }\textbf {\bibinfo {volume} {63}},\ \bibinfo {pages} {022301} (\bibinfo {year} {2001})}\BibitemShut {NoStop}%
\bibitem [{\citenamefont {Bennett}\ \emph {et~al.}(1996)\citenamefont {Bennett}, \citenamefont {Bernstein}, \citenamefont {Popescu},\ and\ \citenamefont {Schumacher}}]{bennett1996concentrating}%
  \BibitemOpen
  \bibfield  {author} {\bibinfo {author} {\bibfnamefont {C.~H.}\ \bibnamefont {Bennett}}, \bibinfo {author} {\bibfnamefont {H.~J.}\ \bibnamefont {Bernstein}}, \bibinfo {author} {\bibfnamefont {S.}~\bibnamefont {Popescu}},\ and\ \bibinfo {author} {\bibfnamefont {B.}~\bibnamefont {Schumacher}},\ }\bibfield  {title} {\bibinfo {title} {Concentrating partial entanglement by local operations},\ }\href {https://doi.org/10.1103/PhysRevA.53.2046} {\bibfield  {journal} {\bibinfo  {journal} {Phys. Rev. A}\ }\textbf {\bibinfo {volume} {53}},\ \bibinfo {pages} {2046} (\bibinfo {year} {1996})}\BibitemShut {NoStop}%
\bibitem [{\citenamefont {Gisin}(1996)}]{gisin1996hidden}%
  \BibitemOpen
  \bibfield  {author} {\bibinfo {author} {\bibfnamefont {N.}~\bibnamefont {Gisin}},\ }\bibfield  {title} {\bibinfo {title} {Hidden quantum nonlocality revealed by local filters},\ }\href@noop {} {\bibfield  {journal} {\bibinfo  {journal} {Physics Letters A}\ }\textbf {\bibinfo {volume} {210}},\ \bibinfo {pages} {151} (\bibinfo {year} {1996})}\BibitemShut {NoStop}%
\bibitem [{\citenamefont {Bose}\ \emph {et~al.}(1999)\citenamefont {Bose}, \citenamefont {Vedral},\ and\ \citenamefont {Knight}}]{bose1999purification}%
  \BibitemOpen
  \bibfield  {author} {\bibinfo {author} {\bibfnamefont {S.}~\bibnamefont {Bose}}, \bibinfo {author} {\bibfnamefont {V.}~\bibnamefont {Vedral}},\ and\ \bibinfo {author} {\bibfnamefont {P.}~\bibnamefont {Knight}},\ }\bibfield  {title} {\bibinfo {title} {Purification via entanglement swapping and conserved entanglement},\ }\href@noop {} {\bibfield  {journal} {\bibinfo  {journal} {Physical Review A}\ }\textbf {\bibinfo {volume} {60}},\ \bibinfo {pages} {194} (\bibinfo {year} {1999})}\BibitemShut {NoStop}%
\bibitem [{\citenamefont {Yang}\ \emph {et~al.}(2020)\citenamefont {Yang}, \citenamefont {Grankin}, \citenamefont {Sieberer}, \citenamefont {Vasilyev},\ and\ \citenamefont {Zoller}}]{yang2020quantum}%
  \BibitemOpen
  \bibfield  {author} {\bibinfo {author} {\bibfnamefont {D.}~\bibnamefont {Yang}}, \bibinfo {author} {\bibfnamefont {A.}~\bibnamefont {Grankin}}, \bibinfo {author} {\bibfnamefont {L.~M.}\ \bibnamefont {Sieberer}}, \bibinfo {author} {\bibfnamefont {D.~V.}\ \bibnamefont {Vasilyev}},\ and\ \bibinfo {author} {\bibfnamefont {P.}~\bibnamefont {Zoller}},\ }\bibfield  {title} {\bibinfo {title} {Quantum non-demolition measurement of a many-body hamiltonian},\ }\href@noop {} {\bibfield  {journal} {\bibinfo  {journal} {Nature communications}\ }\textbf {\bibinfo {volume} {11}},\ \bibinfo {pages} {775} (\bibinfo {year} {2020})}\BibitemShut {NoStop}%
\end{thebibliography}%

\end{document}